\documentclass[
acmsmall,
screen,
authorversion=true,
nonacm=true,
]{acmart}

\usepackage{xcolor}
\usepackage{hyperref}

\definecolor{red}  {rgb}{0.9,0.0,0.0}
\definecolor{green}{rgb}{0.0,0.7,0.0}
\definecolor{blue} {rgb}{0.0,0.0,0.9}

\usepackage{amsfonts}
\usepackage{amsmath,mathtools,commath,nicefrac}

\usepackage{graphicx}
\usepackage{amsthm}

\usepackage{tikz}
\usepackage{gnuplot-lua-tikz}
\usetikzlibrary{shapes,decorations,shadows,positioning,chains,fit,shapes,calc,matrix,backgrounds,arrows}
\usepackage{pgfplots}

\pgfplotsset{compat=newest}
\usepgfplotslibrary{groupplots}
\usetikzlibrary{svg.path}

\usepackage{comment}
\newtheorem{theorem}{Theorem}%[section]
\newtheorem{lemma}{Lemma}%
\newtheorem{corollary}{Corollary}%[section]
\newtheorem{definition}{Definition}%[section]
\newtheorem{problem}{Problem}%[section]
\newtheorem{cnstr}{\textbf{Construction}}%$\!$}

\theoremstyle{definition} % no italics for the newtheorems defined below:
\newtheorem{example}{Example}

\newcommand{\LFFZ}[2][n]{$\del{#1,#2}$-LFFZ}
\newcommand{\decodable }[1][d]{$#1$-decodable}
\newcommand{\decodablek}[2]{$\del{#1,#2}$-decodable}
\newcommand{\minm}[2]{m^*\del{#1,#2}}   % any k
\newcommand{\minmk}[3]{m^*\del{#1,#2,#3}}   % k-weighted codewords
\newcommand{\recresult}{\ceil{\frac{3}{\log_2{3}}\log_2{n}}}

\newcommand{\Code}{\mathcal{C}}
\newcommand{\N}{\mathbb{N}}%
\DeclareMathOperator*{\argmax}{arg\,max}

\newcommand{\ceil}[1]{\left\lceil#1\right\rceil}
\newcommand{\floor}[1]{\left\lfloor#1\right\rfloor}

\newcommand{\Rows}[1]{R\del{#1}}
\newcommand{\Cols}[1]{C\del{#1}}

\usepackage[obeyclassoptions,mode=buildnew]{standalone}

\begin{document}

\title{Invertible Bloom Lookup Tables with Listing Guarantees}

\author{Avi Mizrahi}
\email{avraham.m@cs.technion.ac.il}
\orcid{0000-0002-5715-724X}

\author{Daniella Bar-Lev}
\email{daniellalev@cs.technion.ac.il}
\orcid{0000-0001-6766-1450}

\author{Eitan Yaakobi}
\email{yaakobi@cs.technion.ac.il}
\orcid{0000-0002-9851-5234}

\author{Ori Rottenstreich}
\email{or@technion.ac.il}
\orcid{0000-0002-4064-1238}

\affiliation{%
  \institution{Technion Israel Institute of Technology}
  \city{Haifa}
  \country{Israel}
}

% \renewcommand{\shortauthors}{Mizrahi, et al.}

\begin{abstract}
The Invertible Bloom Lookup Table (IBLT) is a probabilistic concise data structure for set representation that supports a listing operation as the recovery of the elements in the represented set. Its applications can be found in network synchronization and traffic monitoring as well as in error-correction codes.
IBLT can list its elements with probability affected by the size of the allocated memory and the size of the represented set, such that it can fail with small probability even for relatively small sets. 
While previous works \emph{only} studied the failure probability of IBLT, this work initiates the worst case analysis of IBLT that guarantees successful listing for all sets of a certain size. The worst case study is important since the failure of IBLT imposes high overhead.  
We describe a novel approach that guarantees successful listing when the set satisfies a tunable upper bound on its size. To allow that, we develop multiple constructions that are based on various coding techniques such as stopping sets and the stopping redundancy of error-correcting codes, Steiner systems, and covering arrays as well as new methodologies we develop. We analyze the sizes of IBLTs with listing guarantees obtained by the various methods as well as their mapping memory consumption. Lastly, we study lower bounds on the achievable sizes of IBLT with listing guarantees and verify the results in the paper by simulations.
\end{abstract}





\maketitle

\section{Introduction}
The Invertible Bloom Lookup Table (IBLT) has received considerable attention as a probabilistic data structure for the representation of dynamic sets, allowing element insertion and removal along with the ability to list the current elements in the set~\cite{IBLT}.
Applications of the IBLT can be found in traffic monitoring for loss detection or measurement as well as in the design of error-correction codes~\cite{li2016lossradar, li2016flowradar, BiffCodes}.
In particular, communication-efficient set reconciliation with IBLTs, over two or more parties, is useful for database synchronization  and  lightweight blockchain  dissemination protocols~\cite{EppsteinGUV11, multiparty, Graphene19}.

The IBLT has several key advantages such as short update time upon element insertion or removal that is independent on the size of the set. 
While the ability to list elements is probabilistic and  can encounter failures, it was shown to be successful with high probability when the ratio between the number of represented elements and the allocated memory is below some threshold~\cite{IBLT}. Namely, in order to succeed with high probability, the number of elements  at listing time  should be proportional to the memory size, while this number can be temporarily far larger between two listing operations. 
Still, the IBLT might fail, with some small probability, to list even a small set, for example two elements that are mapped to the same $k$ entries.  
 
%This probability for the listing to fail was analyzed extensively as a function of the set size and the amount of allocated memory.  Yugawa et al.~\cite{yugawa2014finite} analyzed the failure probability of an IBLT and suggested a method to improve the success rate using tailored hash functions which guarantee to eliminate listing failure when two elements are stored in the IBLT. Kubjas et al.~\cite{kubjas2020failure} studied the probability for partial listing of the IBLT content and showed a protocol for the set reconciliation that utilizes also partial listings. 

Failing to list the IBLT items results with high overhead cost. For example, in  synchronization among two similar files~\cite{EppsteinGUV11} or between a pair of transaction pools in blockchain networks~\cite{Graphene19}, an IBLT listing failure requires sending a complete file or the full list of transactions to the other end although it is often familiar with a very large portion of such data. In the detection of lost packets in a stream between two switches, a failure to detect missing packets might necessitate retransmission of the complete stream~\cite{li2016lossradar}. Thus, eliminating the listing failure under some conditions can be important to achieve low network operation costs.

To our knowledge \emph{there is no solution that completely prevents listing failures for a predefined domain}. 
In this paper we study the requirements for an IBLT with a listing failure free zone and suggest practical implementations.
Towards such a property, it is easy to see that some restrictions must be assumed. The first refers to the universe from which elements can appear. Its size must be finite and restricted, otherwise within an allocated amount of memory there are two elements with an identical representation. In such a case, listing the set from the IBLT cannot be done accurately even when the set is a singleton, representing one of the pair of elements with the same representation. A second required restriction for guaranteed listing refers to the maximal set size.  This can be helpful for a distinction between sets that allow listing even for a relatively large universe size. Note that this does not restrict the set size from temporarily being large but guarantees a successful listing only when the upper bound on the size is satisfied. Within a domain for the universe size and restricted set, we show that by choosing carefully-designed mapping functions from the universe to the IBLT cells, we can ensure no listing failures.  

\begin{figure}
\centering
\begin{tikzpicture}

\begin{groupplot}[
group style={group size=1 by 2,
             vertical sep=55pt,
},
xlabel={Number of elements in the IBLT},
ylabel={Listing success prob.},
% ylabel style={yshift=-5mm},
% ymin=0,
% ymax=1,
% ymode=log,
% xmin=1,
% xmax=7,
% max space between ticks=18,
ymajorgrids,
% yminorgrids,
% xmajorgrids,
% xminorgrids,
% height=.3\textwidth,
% width=.65\textwidth,
height=.3\textwidth,
width=.9\textwidth,
x grid style={lightgray!92!black},y grid style={lightgray!92!black},
legend cell align={left},
legend pos=south west,
]

\nextgroupplot[
title={table length $m=15$, universe size $n=|U|=25$},
xmax=8,
ytick = {0.6,0.7,0.8,0.9, 1}
]
\addplot[color=red, mark=o] table [x=N, y={m=15_k=3_MM3_iblt}, col sep=comma]%
        {data/dffz_vs_iblt_d3_m15_n25.csv};%
\addlegendentry{IBLT $m$=15, $k$=3}

\addplot[color=brown, mark=square] table [x=N, y={m=15_d=3_ols}, col sep=comma]%
        {data/dffz_vs_iblt_d3_m15_n25.csv};%
\addlegendentry{LFFZ, OLS $m$=15, $k$=3, $d$=3}

\addplot[color=blue, mark=diamond] table [x=N, y={m=15_d=3_k=3_rec}, col sep=comma]%
        {data/dffz_vs_iblt_d3_m15_n25.csv};%
\addlegendentry{LFFZ, Construction C $m$=15, $k$=3, $d$=3}

\nextgroupplot[
title={table length $m=64$, universe size $n = |U|=381$},
ylabel={},
xmax=6,
xtick={1,...,6},
ymin = 0.99918, 
ymax = 1.0001,  
ytick = {0.9992, 0.9994, 0.9996, 0.9998, 1}, 
yticklabel style={
    /pgf/number format/precision=4
},
]
\addplot[color=red, mark=o] table [x=N, y={m=64_k=4_MM3_iblt}, col sep=comma]%
        {data/dffz_vs_iblt_d5_m64_n381.csv};%
\addlegendentry{IBLT $m$=64, $k$=4}

\addplot[color=orange, mark=triangle] table [x=N, y={m=64_d=5_rec}, col sep=comma]%
        {data/dffz_vs_iblt_d5_m64_n381.csv};%
\addlegendentry{LFFZ, Construction A $m$=64, $d$=5}

\addplot[color=blue, mark=diamond] table [x=N, y={m=64_d=3_k=4_rec}, col sep=comma]%
        {data/dffz_vs_iblt_d5_m64_n381.csv};%
\addlegendentry{LFFZ, Construction C $m$=64, $k$=4, $d$=3}

\end{groupplot}

\end{tikzpicture}
\caption{\label{fig:intro}
The listing success probability of the traditional IBLT vs. the suggested IBLT with a listing failure free zone (LFFZ).
The probability is a function of the table length and number of elements in the IBLT.
In the traditional IBLT, listing can fail even with a set of two elements, while for the IBLT with LFFZ listing is always successful for any number of elements up to some known parameter $d$, allowed by the table length and the universe size. 
}
\end{figure}
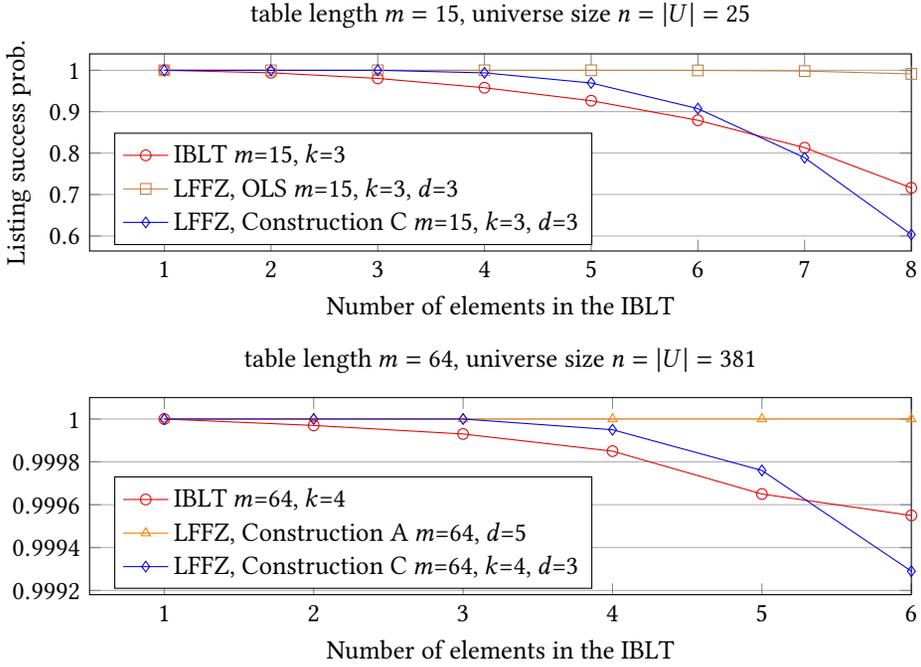

Our work is the first to allow IBLTs with the following property: listing elements always succeeds for any set of up to $d$ elements from a finite universe $U$ of size $n$.
\autoref{fig:intro} illustrates the listing success probability as a function of the number of elements in the table. For the traditional IBLT and the IBLT with listing failure free zone (LFFZ), the guarantee we present to avoid listing failures for all sets of size at most $d$.
In the left figure, for a table length $m=15$, elements drawn from a universe of size $n=25$, while a failure can occur in the IBLT even with two elements, these are avoided for sets up to size $d=3$ in the IBLTs with LFFZ.
In the right figure,
for a table length $m=64$ with a universe of size $n=381$, failures can be avoided in the IBLT with LFFZ for sets up to sizes $d=3,5$.
When the number of IBLT cells each element is mapped to is fixed, it is indicated as $k$.

As will be explained in detail in the paper, an IBLT can be represented by its $m\times n$ \emph{mapping matrix} which indicates for each element the cells it is mapped to. There is a strong connection between the mapping matrix of an IBLT and the parity-check matrix of error-correcting codes. Listing any set of $d$ elements from the universe is successful if any set of at most $d$ columns in the mapping matrix has at least one row of weight one. This property is very similar to the \emph{stopping redundancy} of error-correcting codes and in fact we observe that if the stopping redundancy of a length-$n$ code with minimum Hamming distance $d+1$ is $\rho$ then there is an IBLT with $m=\rho$ cells that can successfully list any $d$ elements. Interestingly, while we can use any known results on the stopping redundancy of codes to construct IBLT with listing guarantees, we also observe that these two properties are not equivalent. For example, in order to successfully list any $d=3$ elements, the best construction based upon stopping redundancy of codes requires $2\lceil \log_2 n \rceil -1$ cells, while we show how to accomplish the equivalent result for IBLTs with only $\ceil{\frac{3}{\log_2{3}}\log_2{n}} \approx \ceil{1.89\log_2{n}}$ cells. The goal in both problems is to find a matrix satisfying the property that every set of at most $d$ columns contains a row of weight 1. However, while the matrices in the stopping redundancy of codes should also be of a small dimension (corresponding to the code redundancy), the mapping matrix of IBLTs does not impose such a constraint and can even be of full row rank.

\paragraph{Our contributions.}
This work is the first to introduce the IBLT with Listing Failure Free Zone (LFFZ).
We relate this property to the minimum size of the stopping sets of the matrix representing the mapping of elements to cells of the IBLT.
Moreover, we present a variety of constructions for matrices allowing IBLTs with LFFZ, both by linking between such matrices to well studied problems, and by introducing an entirely new family of recursive constructions that are suitable for any set of parameters.
Furthermore, to assess the presented constructions, we provide theoretical analysis of lower bounds and upper bounds on the size of such matrices.
Finally, we implement the constructions for studying the dependency of their parameters for practical values.
A summary of the constructions and their parameters is given in \autoref{table:results} and an extensive summary of the contributions in the paper appears in \autoref{sec: summary}.

\section{Terminology and problem statement}\label{sec: problem definition}
In this section, we formally define guarantees on the listing performance of the IBLT.
We start with a background on the Bloom filter and the IBLT.
Then we give preliminary definitions and define the IBLT with Listing Failure Free Zone (LFFZ) and the optimization problems we cover.
Lastly, we show the basic properties of the LFFZ before we present constructions in the next sections.
\autoref{table:notations} summarizes the notations used throughout the paper.

\subsection{Probabilistic Data Structures}\label{subsec: data structures}
The IBLT can be viewed as a natural extension of the Bloom filter. The Bloom filter is also used for set representation and answering membership queries in a probabilistic accuracy, but does not support listing of its elements.

\begin{figure}
\centering
\begin{minipage}{.43\textwidth}
    \centering
    \includegraphics[
    trim=0.0cm 0.0cm 0.0cm 1.5cm,
    width=.93\linewidth
    ]{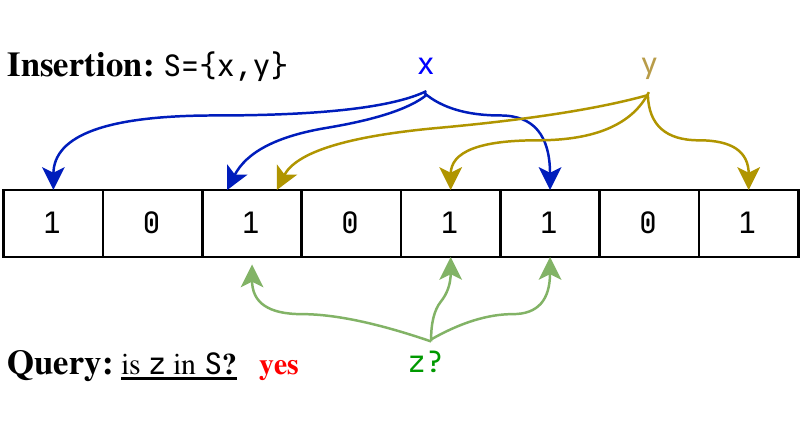}
    \captionof{figure}{A Bloom filter representing a set $S = \{x,y\}$, falsely indicating the membership of an element $z \notin S$.}
    \label{fig:BF}
\end{minipage}\hfill
\begin{minipage}{.55\textwidth}
    \centering
    \includegraphics[
    % trim=0cm 0.8cm 0cm 0.5cm,
    width=.85\textwidth
    ]{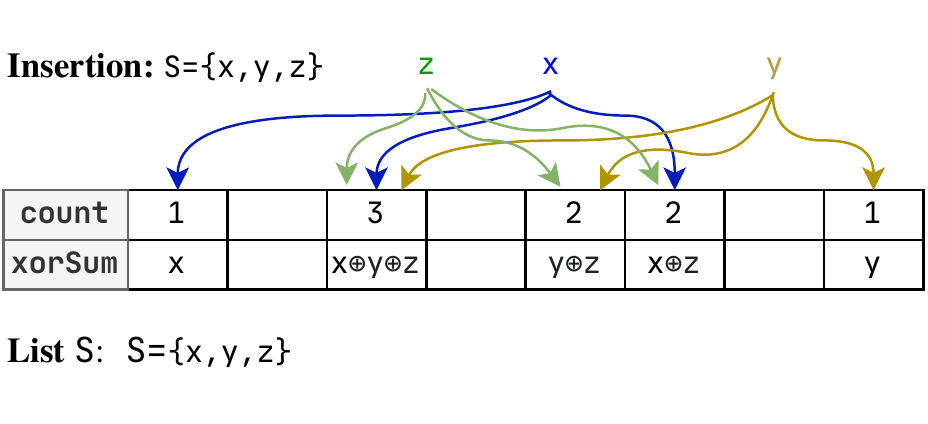}
    \captionof{figure}{An IBLT representing a set $S=\{x,y,z\}$ with a similar mapping of elements as in~\autoref{fig:BF}.
    Element listing is possible starting from the most left cell with \texttt{count=1}, identifying $x$ as a member of $S$ and removing it from the other two cells. Then both $y$ and $z$ can be listed by the cells that become pure.
    }
    \label{fig:IBLT}
\end{minipage}
\end{figure}

\paragraph{Bloom filter~\cite{Bloom}.} A popular  probabilistic data structure used for set representation, supporting element insertion and answering membership queries. There can be two kinds of errors in membership queries: a false positive (when an element $x \notin S$ is reported as a member of a represented set $S$) and a  false negative (when an element $x \in S$ is reported as a nonmember of $S$). The Bloom filter encounters false positives and has no false negatives. It is built as an array of bits, where hash functions are used to map elements to locations in the array. With initial values of zero bits, the elements of $S$ are first inserted to the filter, setting all bits $h_1(x),\ldots,h_k(x)$ pointed by $k$ hash functions $h_1,\ldots,h_k$. Upon a query, the $k$ bits mapped by the queried element are examined and a positive answer is returned only when the bits are all set.  The Bloom filter is illustrated in \autoref{fig:BF}.  The Bloom filter cannot support removal (deletions) of elements from the represented set. This is since resetting bits to zeros results in forbidden false negatives of all remaining elements that map to these bits. 

\paragraph{Invertible Bloom Lookup Table (IBLT), Invertible Bloom filter (IBF)~\cite{IBLT, EppsteinGUV11}.}
IBLTs and IBFs
generalize Bloom filter functionality in the representation of a set $S$. In addition to  element insertion, deletion, and membership queries, they also support  a complete listing of the elements with high probability when the number of represented elements is at most some predefined threshold implied by the allocated amount of memory. Following their similarity, in particular in the listing process, for the simplicity of presentation, we refer to the basic form of them\footnote{Typically the IBF is used to keep a set of items while the IBLT supports storing key-value pairs. In some cases, they contain a third field in a cell. As their listing process is based on the counter field in a cell our results are applicable to both data structures.}.
An IBLT is illustrated in \autoref{fig:IBLT}.
Assume that each element is associated with a (unique) key. The IBLT keeps an array of  cells, each with two fields:
\begin{enumerate}
    \item count - stores the number of elements mapped to the cell;
    \item xorSum - stores the cumulative XOR (exclusive or) of the keys mapped to the cell.
\end{enumerate}
Each element is mapped to several cells, e.g., with hash functions. Upon insertion the counter in a cell is incremented and the xorSum field is XORed with the key. Deletion is supported by decrementing the counter and updating the xorSum as in insertion. We refer to a \emph{pure cell} as a cell with a single inserted element. Such a cell can be identified based on the count field. The \emph{listing procedure} of the IBLT uses a \emph{peeling} process that repeatedly looks for a pure cell. Whenever such a cell is found, an element is identified based on the keySum field which is necessarily the key of the element. Then, we can remove the element from the other cells it is mapped to, thus reducing the number of remaining elements in these cells. The process fails if no pure cells can be found at some step, earlier to the listing of all elements.
Listing is successful with high probability when the ratio between the number of represented elements to the allocated memory is below some threshold (see \autoref{subsec: IBLT failure probability}). 
We say that an IBLT is \emph{regular} if each element is mapped to a fixed number of cells, and otherwise we refer to it as irregular.

While Bloom filters can refer to the representation of sets taken from finite or infinite universes, in the context of IBLT, typically finite universes are assumed. This follows the encoding of elements uniquely within the IBLT cells (in particular in the xorSum field). We follow this assumption and refer to the IBLTs representing sets as subsets of some known \emph{finite universe}. 

\subsection{Problem Statement}\label{sec: problem statement}
We begin with the definition of the IBLT with LFFZ that guarantees the success of its listing operation for sets of a restricted size.%when the size of the represented set is restricted. 
\begin{definition}[\LFFZ{d}]
\label{def:lffz}
The \textbf{$\del{n,d}$-Listing Failure Free Zone} (\textbf{LFFZ}) of an IBLT allows a finite universe $U=\{1,\dots,n\}$ of size $n$ and a maximal set size $d$, if for any IBLT for a set  $S\subseteq U$ satisfying $|S|\leq d$, the list method returns exactly the complete set $S$.
\end{definition}

For a finite universe of elements, we can represent the mapping of elements to cells of the IBLT as a binary matrix. 
In such a matrix, columns represent elements from the universe and rows are matched to the IBLT cells. The matrix size is $m \times n$ where $m$ is the table length (number of cells) and $n = |U|$ refers to the universe size. 

The mapping is implied by the $k$ hash functions of the table such that each element is mapped into $k$ cells. To construct an IBLT with LFFZ, we want to replace the general hash functions with carefully designed functions to map elements to cells.
By doing so wisely, we can assure listing for any set of elements up to a given size.

\begin{definition}[Mapping Matrix]\label{def: mapping matrix}
An $m \times n$ binary matrix $M$ defines an IBLT with $m$ cells over a finite universe $U = \cbr{1, \ldots, n}$ in the following way.
The element $i \in U$ is mapped to the $j$-th cell if and only if $M_{j,i}=1$.
Each element is mapped to at least one cell such that $\sum_{j} M_{j,i} > 0$ for all $i \in U$.
\end{definition}

%The requirement for no zero-columns is to prevent elements from not being mapped into cells.

Recall that each cell of the IBLT has two fields: count and xorSum. Each counter simply indicates the number of elements in the set mapped to the cell. We refer to the $m$ counter values simply as the \emph{counter array}.
For an IBLT with a mapping matrix $M$ that stores a set $S \subseteq U$, the values of the counter array equal the sum of the columns in $M$ that correspond to the elements of $S$.

\begin{figure}[h!]
\centering
\begin{minipage}{.65\textwidth}
\begin{example}\label{example:matrix}
    For the universe $U = \{1, \ldots, n=6\}$ consider the binary matrix $M$ on the right.
    An IBLT based on $M$ has $m=5$ cells, each associated with a row of $M$.
    Such an IBLT, when containing for instance the set $S=\cbr{1, 3, 4}$ (such that $S \subseteq U$), has the counter array $\del{2, 1, 2, 0, 1}$ as the sum of the entries in the first, third, and fourth columns.
\end{example}
\end{minipage}\hspace{.5cm}%\hfill
\begin{minipage}{.25\textwidth}
\vspace{-.3cm}
    \[
    % this matrix is with best achievable $d$ for $m=5,n=6$ for both LFFZ and FPPZ, with $d=3,d=1$ respectively
    M=
    \begin{bmatrix}
    1 & 1 & 1 & 0 & 0 & 0\\
    0 & 0 & 0 & 1 & 1 & 1\\
    1 & 0 & 0 & 1 & 0 & 0\\
    0 & 1 & 0 & 0 & 1 & 0\\
    0 & 0 & 1 & 0 & 0 & 1\\
    \end{bmatrix}
    \]
\end{minipage}
\end{figure}

For an IBLT to list successfully, at each peeling iteration we must have at least one pure cell. %Namely, the counter array must have at least one counter that equals one at each iteration. 
To be able to list successfully any set of size $|S| \le d$, the counter array that corresponds to any such set should have at least one counter with the value 1. These IBLT guarantees can also be expressed as properties of its binary mapping matrix $M$. 
In particular, it relates to \emph{stopping sets}~\cite{stoppingSets}, which is a combinatorial structure that was defined in terms of matrices for the decoding procedure of error-correcting codes~\cite{schwartz2006ss}:

\begin{definition}[Stopping Set]\label{def: stopping set}
%A stopping set in a matrix $M$ is a set of columns with the property that the sub-matrix these columns form does not contain a row of weight one.
Let $M$ be a matrix and $S$ a non empty set of its columns.
The weight of a row in the sub-matrix implied by $S$ is defined as the the number of non-zero coordinates in it.
% Given a matrix $M$ and a subset of its columns, we refer to its implied sub-matrix and to the number of non-zero coordinates in its rows as their weight.
% A stopping set in $M$ is a subset of columns without a row of weight one.
$S$ is called a stopping set if it has no row of weight one.
%Where the weight of a vector is defined as the number of non-zero coordinates in it.
The stopping distance of $M$, denoted by $s(M)$, is the size of the smallest stopping set in $M$.
\end{definition}

Consider the matrix $M$ and the set $S=\cbr{1, 3, 4}$ from \autoref{example:matrix}. The set is not a stopping set as can be deduced from the second and fifth counters having the value of 1, which corresponds to rows of weight 1 (second and fifth) in the sub-matrix of $S$. However, the set $S \cup \cbr{6} = \cbr{1, 3, 4, 6}$ is a stopping set since its counter array $\del{2,2,2,0,2}$ has no value of 1.
It is possible to verify that the stopping distance of $M$ satisfies $s(M)= 4$ as there is no stopping set of size 3. In the peeling process of the set $S$, the second counter indicates a pure cell.
The value of the xorSum field in that cell allows us to derive the single element in $S$ mapped to that cell, which is the element $4 \in U$. We can then remove it from the IBLT's cells according to the fourth column in the matrix and continue the peeling process with the set $\{1,3\}$ and the counter array $\del{2, 0, 1, 0, 1}$.

To ensure that the peeling process succeeds for any set of at most $d$ elements from $U$, we require that $M$ does not contain any stopping set of size $d$.
In other words, we are looking for a matrix $M$ with a stopping distance strictly greater than $d$.

\begin{definition}[\decodable{} matrix]\label{def:d-decodable}
An $m\times n$ binary matrix $M$ is called \decodable{} if its stopping distance is at least $d+1$, i.e., $s(M)\geq d+1$.
\end{definition}

Note that $M$ from \autoref{example:matrix} is \decodable[3], as any set of $d=3$ columns is not a stopping set.
Combining \autoref{def:d-decodable} with \autoref{def:lffz} results with the following statement.

\begin{corollary}
Any $m \times n$ \decodable{} matrix is equivalent to an IBLT with $m$ cells and \LFFZ{d}.
\end{corollary}

From this point on we focus our discussion on the exploration of \decodable{} matrices, which by the latter corollary are equivalent to IBLT with LFFZ.
Naturally, we seek to study the effect of the universe size $n$ and threshold value $d$ on the memory used by the table $m$, and find matrices that minimize it. This problem is formalized as follows.
\begin{problem}\label{problem:nd}
Given $n,d\in \N$, $d \leq n$ find the value:
\[\minm{n}{d} = \min{\cbr{m : \exists M \in \cbr{0,1}^{m \times n}, M \text{ is \decodable{}}}}\] and a corresponding \decodable{} $\minm{n}{d} \times n$ matrix.
\end{problem}

We might also require equally weighted columns of $M$, with a small weight as possible.
This property ensures a small fixed number of read or write memory accesses when querying, inserting or removing elements from the IBLT.
We study IBLTs with columns of equal weight, also referred as regular IBLTs, as this is the family of IBLTs that has been mostly studied in the literature.
As in~\autoref{def:d-decodable}, we use the notation of \emph{\decodablek{d}{k} matrix} to refer to a \decodable{} matrix such that each of its columns is of weight $k$.
Similarly to \autoref{problem:nd}, we are interested in \decodablek{d}{k} matrices having the minimal number of rows.
\begin{problem}\label{problem:ndk}
Given $n,d,k\in \N$, $d \leq n$ find the value:
\[\minmk{n}{d}{k} \hspace{-0.3ex}= \hspace{-0.3ex}\min \hspace{-0.3ex} {\cbr{m \hspace{-0.3ex}: \exists M \hspace{-0.3ex}\in\hspace{-0.3ex} \cbr{0,1}^{m \times n}, M \text{ is \decodablek{d}{k}}}}\] and a corresponding \decodablek{d}{k} $\minmk{n}{d}{k} \times n$ matrix.
\end{problem}

\begin{table}
\centering
\begin{tabular}{clccccl}
\hline
Methodology & Code & & $d$ & $m$ & $k$ \\

\hline 
Bloom filters & EGH Bloom filter~\cite{EGH} & \autoref{cor: LFFZ from FPFZ} & any & $O\del{d^2\log_2{n}}$ & $O\del{d\log_2{n}}$ \\
(Section~\ref{section_bloom_constructions}) & OLS Bloom filter~\cite{FPFZ} & \autoref{cor: LFFZ from FPFZ} & any & $d\sqrt{n}$ & $d$ \\ 
\hline 
Linear codes & Extended Hamming & \autoref{cor:from linear} & $3$ & $2\log_2{n}-1$ & - \\
(Section~\ref{section_linear_codes_constructions}) & Simplex~\cite{etzion2006stopping} & \autoref{cor:from linear} & $\frac{n-1}{2}$ & $n - \log_2\del{n+1}$ & - \\
% 1st order Reed-Muller~\cite{etzion2006stopping} & \autoref{cor:from linear} & $n/2$ & $O\del{\frac{1}{3}n\log{n}}$ & & \\
& Array codes~\cite{Esmaeili09} & \autoref{cor:from linear} & $2k-1$ & $k\ceil{\sqrt{2n}}$ & $2\le k\le 4$ \\
& LS-LDPC~\cite{laendner2007ldpc} & \autoref{cor:from linear} & $5$ & $6\ceil{\sqrt{n/6}} + 3$ & $3$ \\
& BCH~\cite{roth_2006} & \autoref{cor: from BCH} & $4$ & $4\ceil{\log\del{n+1}}$ & $2\ceil{\log\del{n+1}}$ \\
% Array-LDPC~\cite{rosnes2014minimum} & \autoref{cor:from linear} & $7-16$ & $q\sqrt{n}$ & $\sqrt{n}$ & for prime $q\leq 79$ \\
\hline
Combinatorial  & Steiner Triple~\cite{colbourn1999triple}
 & \autoref{cor: Steiner and IBLT} & 3 & $\sqrt{6n} + O(1)$ & 3 \\
structures & Steiner 3-design~\cite{blanchard1995construction} & \autoref{cor: Steiner and IBLT} & $\ceil{\frac{q+1}{2}}$ & $O\del{d^2\sqrt[3]{n}}$ & $q+1$ \\  
(Sections~\ref{section_steiner_constructions}-\ref{section_covering_arrays_constructions})  & Covering Arrays~\cite{CoveringArrays} & \autoref{the: upper bound covering arrays} & any & $ \frac{d\log_2(n)}{\log_2\left(2^d / (2^d-1)\right)}$ & - \\
 & Covering Codes & \autoref{theorem_lower_bound_k=2_d=3} & 3 & $\ceil{2\sqrt{n}}$ & 2 \\
\hline
Recursive                                       & \autoref{const: improved d-decodable}  & \autoref{th: recursive improved}     & $O(1)$ & {$O\del{\log_2^{\floor{\log_2 d}}{n}}$} & - \\
constructions                                   & \autoref{const: improved 3-decodable}  & \autoref{th:d3construction}          & $3$    & $\recresult$ & - \\
(\autoref{section_recursive_constructions}) & \autoref{const: basic (d,k)-decodable} & \autoref{th: recursive d,k improved} & any    & see \autoref{th: recursive d,k improved} & any \\
\hline
\end{tabular}
\caption{Constructions overview based on existing and suggested codes}
\label{table:results}
\end{table}

\subsection{Early Observations}\label{subsec: basic properties}

The trivial case of $d=1$ implies no additional requirements for $M$; in particular, the matrix with only one row of ones meets them. Thus, $\minm{n}{1}=1$ and $\minmk{n}{1}{k} = k$ for any integers $n,k$.
Note that if $M$ has two identical columns, these two columns form a stopping set of size two and hence $M$ is not \decodable[2].
Accordingly, from now on we assume that there are no identical columns in $M$.
Next, we show that the case of $d=2$ is also easily solved.

\begin{theorem}\label{th:pre}
For any integers $n, k$, the following holds,
\begin{enumerate}
    \item $\minm{n}{2}=\ceil{\log_2\del{n+1}}$.
    \item $\minmk{n}{2}{k} = \min\cbr{m\in \N : \binom{m}{k}\ge n}$.
\end{enumerate}
\end{theorem}
\begin{proof}
Any binary ${m \times n}$ matrix with unique columns is \decodable[2]. Since every two columns are distinct, they must differ at some row, which is necessarily a row of weight one in the corresponding ${m \times 2}$ sub-matrix.
Then, the number of columns $n$ is bounded by all possible non-zero columns and all possible non-zero columns of weight $k$, respectively.
\end{proof}

Lastly, we state simple results for the case where $d \ge 3$, the proof can be found in \autoref{app:A}.
\begin{theorem}\label{th:basic}
For any integers $n,k$ and $d\ge 3$ we have that
\begin{enumerate}
    \item $m^*(n,d,k) \ge m^*(n,d)\ge \max\{d,\ceil{\log_2\del{n+1}}\}$ and \\ ${m^*(n,d,k)\ge \max\{d,k,\ceil{\log_2\del{n+1}}\}}$.
    \item $m^*(n,d) \ge m^*(n,d-1)$.
    \item $m^*(n,n) = n$ and for $d<n$, $m^*(n,d)\leq n-1$.
    \item $m^*(n,d,k) \leq n+k-1$ and $m^*(n,d,k=1) = n$.
    \item $m^*(n,d,k+1)\leq m^*(n,d,k) +1$.  
\end{enumerate}
\end{theorem}

\subsection{Summary of the Results}\label{sec: summary}
In this work, we introduce the IBLT with Listing Failure Free Zone (LFFZ), an IBLT that guarantees a successful listing for any set of size at most $d$ from a finite universe of size $n$, where $n,d$ are given parameters. The main contributions of this paper are as follows.
\begin{itemize}
    \item To the best of our knowledge, we are the first to provide IBLT with guaranteed listing  whenever the number of elements is at most some given threshold $d$. We define a \decodable[d] matrix of size $m\times n$ to be a matrix that has no stopping set of size $d$ or less and show that any such a matrix is equivalent to an IBLT with $(n,d)$-LFFZ and $m$ cells. We also consider \decodable[(d,k)] matrices which are \decodable[d] matrices with a fixed column weight $k$ and ensure a fixed (small) number of read and write operations when querying.
    \item We present a variety of constructions for \decodable[d] and \decodable[(d,k)] matrices.
    As described next, some of the constructions are obtained by connecting the above LFFZ requirements to a variety of known problems in coding theory and design theory, while others are developed entirely by us and can be applied for a generic set of parameters.
    \begin{itemize}
        \item We identify 
        other problems which are related to stopping sets: Bloom filters with FPFZ, stopping redundancy of linear codes, Steiner systems, and covering arrays. We conduct a study of constructions which we identify as allowing  \decodable[d], \decodable[(d,k)] matrices with a small number of rows. 
        These constructions provide upper bounds on the values of $m^*(n,d)$ and $m^*(n,d,k)$.    
        \item We design new families of recursive constructions for \decodable[d] and \decodable[(d,k)] matrices, which are suitable for all parameters $n,d$ and $k$. These constructions, in particular, solve infinite sets of parameters $n,d,$ and $k$ which have not been solved so far. 
        In addition, we prove recursive upper bounds on $m^*(n,d)$, $m^*(n,d,k)$ and analyze them for $m^*(n,d)$ to obtain a closed form expression. We show that there are parameters for which our recursive construction outperforms any other construction (See \autoref{fig: lffz size}).
    \end{itemize}
    A detailed summary of all the constructions and their relevant parameters is given in \autoref{table:results}. 

    \item We provide theoretical analysis of several lower bounds on $m^*(n,d)$ and $m^*(n,d,k)$. These are lower bounds on the minimum number of rows of any \decodable[d] and \decodable[(d,k)] matrices, independently of any construction. 
    \item We examine the dependency of the parameters $m, n, d, k$ in the various constructions for practical values.
\end{itemize}

\section{Related Work}
In this section we give known properties of the IBLT, its applications, and related work.

\subsection{IBLT Known Properties and Related Approaches}
\subsubsection{IBLT failure probability analysis}\label{subsec: IBLT failure probability}
IBLT's ability to list the elements of the set is only probabilistic and can potentially fail also for small sets. Consider an IBLT with particular parameters (memory size (number of cells) $m$ and number of hash functions $k$) such that each hash function is distributed uniformly over the cells and are mutually independent. There exists some threshold $t$ on the size $|S|$ of the represented set $S$ such that when $|S| < t$ the listing operation succeeds with high probability of $1 - o(1)$. On the other hand, if $|S| > t$ listing succeeds with probability $o(1)$.  
Denote $c_k^{-1} = sup \{\alpha: 0 < \alpha < 1: \forall x \in (0,1), 1- e^{- k \alpha x^{k-1}}< x \}$. For instance $c_k = 1.222, 1.295, 1.425, 1.570, 1.721$ for $k=3,4,5,6,7$, respectively. The IBLT satisfies that if $m > (c_k + \epsilon) t$ for some $\epsilon > 0$ then listing fails with probability $O(t^{-k+2})$ whenever the  set size $|S|$ satisfies $|S| \le t$.
IBLT listing can potentially fail also when representing a set $S$ of size two. In case both elements in the set map exactly to the same $k$ cells, none of the cells include a single element so listing of the set is impossible.

In~\cite{yugawa2014finite}, the authors provided an upper bound on the listing failure probability of the IBLT. This analysis is based upon an enumeration of the number of the so-called stopping sets of a given size when the IBLT can be represented by a matrix  that is divided into blocks and the weight of every column in each block is one. This result was then extended in~\cite{kubjas2020failure} to study the success probability of partial listing of the IBLT, that is, retrieving at least only a fraction of the stored elements in the IBLT. A recent work~\cite{Lazaro_irregular} extended the results from~\cite{IBLT}, that studied the asymptotic performance of regular IBLTs, to irregular IBLTs. By analyzing and deriving the density evolution they were able to obtain the load threshold that guarantees successful listing with high probability. 

\subsubsection{Efficient IBLT listing}
As detailed above, for listing its elements, the IBLT  uses a peeling process that repeatedly looks for a pure cell. When such a cell is found, an element can be added to the list based on the keySum field and should be removed from all other cells it is mapped to. The process fails if no pure cells can be found at some step, earlier to the listing of all elements.
As discussed in \autoref{sec: problem statement}, an IBLT can be represented by its binary mapping matrix and such a matrix will have strong connection to parity check matrices of error-correcting codes.
The requirement on the mapping matrix to guarantee successful listing of any $d$ elements is that every set of at most some $d$ columns in the matrix has a row of weight one. When studying parity check matrices of error-correcting codes, the same property is considered in order to determine the success of iterative decoding over the binary erasure channel. While for error-correcting codes it may be possible to decode the message even when this property does not hold (since it is primarily used for efficient decoding), for IBLTs this property guarantees both efficient and successful listing. The question of determining the minimum requirements on the mapping matrix such that listing is successful, and not necessarily by the peeling procedure, is interesting by itself that is left for future work and is out of the scope of this paper.

 In our constructions for \autoref{problem:ndk}, e.g. the ones by Steiner systems (Section~\ref{section_steiner_constructions}), the weight of every column in the mapping matrix is fixed. This not only guarantees the same low complexity for the insertion and deletion operations, but also significantly reduces the listing complexity since every peeling step requires a small fixed number of operations. A further improvement of the listing procedure by peeling is discussed in~\cite{jiang2017parallel}. Here, the goal is to efficiently find the so-called \emph{$t$-core set} of vertices in a hypergraph by repeatedly removing in parallel all vertices (and their neighborhoods) of degree less than $t$, while IBLT corresponds to $t=2$. The authors studied this problem for arbitrary $t$ and regular hypergraphs to analyze the number of rounds of this parallel peeling process and to find conditions for the $t$-core set, which imply successful listing for IBLT, to be empty. Finding conditions on the mapping matrix such that listing is successful by parallel peeling in a given number of rounds is another interesting problem which is left for future work. 

An interesting property for some constructions such as those based on Bloom filters (Section~\ref{section_bloom_constructions}), allows also a simplified listing process. In these two constructions, for any set $S$ satisfying $|S| < d$, each element in $S$ is mapped to at least one cell that contain no other elements. This allows listing each element independently from the IBLT, without the need to remove from it some other previously-listed elements.

\subsection{Networking Applications of IBLT}
We present major network applications of IBLT. In each of these domains, a failure to list the elements of the IBLT has high cost overhead. In network monitoring it can result in missing flows of high traffic and the inability to identify lost packets. For set reconciliation, it can require sending the complete very large files or transaction pools, drastically increasing communication cost. In error-correction codes, a message might have to be retransmitted even in some cases it was received with small amount of errors.

\subsubsection{Set reconciliation}\label{subsec: set reconciliation}
Assume two entities $A,B$, each holding a set $S_A,S_B$, respectively.
Set reconciliation protocols enable $A$ and $B$ to synchronize their sets, i.e., to find the union of the two sets $S_A \cup S_B$, with efficient communication cost. Numerous network applications use set reconciliations, for example: in blockchain systems the sets can be transaction pools that are synchronized frequently~\cite{Graphene19}; in Peer-to-Peer (P2P) systems the sets are file blocks.

Trivially, each entity can send its set as a list, but a lot of communication can be saved, especially when the two sets are large and have many common elements. In these cases, IBLTs can dramatically improve the communication cost, by sending data whose size is linear in the symmetric difference $\Delta = \del{S_A \setminus S_B} \cup \del{S_B \setminus S_A}$.

IBLT-based reconciliation protocols were suggested by~\cite{EppsteinGUV11,IBLT,Graphene19}.
We now describe their main idea, with the assumptions that $S_B \subseteq S_A$, implying $\Delta = \del{S_A \setminus S_B}$ and only $B$ should reconcile $S_A$;
for full details in the general case the reader is referred to the above works.
$A$ constructs an IBLT with $S_A$ and sends it to $B$, then $B$ removes $S_B$ from the IBLT.
Now the IBLT contains $S_A \setminus S_B$, which are the elements that were inserted by $A$ and were not removed by $B$. %, and $S_B \setminus S_A$, the elements that $B$ removed that were not inserted by $A$.
% It is therefore possible to have a negative cell count, and in particular, a cell with a count of $-1$ that contains a single element from $S_B \setminus S_A$.
%Such cell is also considered as a pure cell in the peeling process, and its element is then \emph{inserted} to the IBLT, as opposed to a pure cell count 1 where the element is removed.
To reconcile the set $S_A$ in $B$, the IBLT should list successfully the set of elements in it, which is $\Delta$. To list $\envert{\Delta}$ elements with high probability, the IBLT that $A$ sends can be configured to use $O\del{\envert{\Delta}}$ cells~\cite{EppsteinGUV11}.
Particularly, the size of the IBLT does not depend on the sizes of $S_A$ and $S_B$, which can be excessively large relative to $\envert{\Delta}$.

\subsubsection{Error-correction codes}\label{subsec: apps: biff codes}
A Biff code~\cite{BiffCodes} is an error-correction code that uses an IBLT as the redundancy of a message. The code is systematic, namely the input message is embedded in the encoded output.
It is designed through a reduction from a set reconciliation method (\autoref{subsec: set reconciliation}), where the sets are the input message and the received message, potentially with errors.
The encoding of a message $X=x_1,\dots,x_n$, is the original message $X$, along with an IBLT $I$ that contains the set of pairs $X'=\cbr{\del{x_1, 1},\dots,\del{x_n, n}}$.
Assume, for the simplicity of the description, that $I$ can be received without errors; the case where $I$ is sent over the same channel as the message such that the IBLT cells can be also erroneous, is also handled in the paper~\cite{BiffCodes}.

Given an erroneous message $Y=y_1,\dots,y_n$ along with the IBLT $I$, the decoder creates the set $Y'=\cbr{\del{y_1, 1},\dots,\del{y_n, n}}$ and reconciles it using $I$.
After a successful reconciliation, the set $X'$ is used to reconstruct the original message $X$. 
Let $\delta$ be the number of errors in the message $Y$ and $i_j$ be the erroneous indices for $j \in \cbr{1, \dots, \delta}$, i.e. $x_{i_j} \ne y_{i_j}$.
The heart of this process includes the IBLT with the set $X'$ inserted and the set $Y'$ removed.
The listing procedure  results, if successful, in the set
$\cbr{\del{x_{i_1}, i_1}, \del{y_{i_1}, i_1}, \del{x_{i_2}, i_2}, \del{y_{i_2}, i_2}, \dots, \del{x_{i_\delta}, i_\delta}, \del{y_{i_\delta}, i_\delta}}$
of size $2\delta$.
If the number of errors $\delta$ can be upper bounded, the size of the IBLT $I$ can be configured such that $2\delta$ items can be listed with high probability.
Therefore, the redundancy of the Biff code is $O\del{\delta}$~\cite{BiffCodes}, the size of $I$.

\subsection{Bloom filter with false positive free zone}
The traditional Bloom filter~\cite{Bloom} supports the representation of a (finite) set $S$ as a subset of an infinite or finite universe $U$. It avoids false negatives in set membership queries but  can encounter false positives. 
Two recent papers~\cite{EGH, FPFZ} suggest Bloom filter constructions with a false-positive-free zone: any filter for set of up to $d$ elements from a finite universe $U$ is guaranteed to have no false positives and no false negatives.
These constructions replaced the general-purpose hash functions of the filter with a carefully-designed mapping of elements to cells.
We thoroughly discuss the relation between this concept and IBLT with LFFZ in \autoref{section_bloom_constructions}.

\begin{table}
\centering
\begin{tabular}{|cl|}  
\hline
Symbol & Meaning \\ 
\hline  
$S$ & represented set\\
$U$ & universe from which elements are selected\\
$n$ & universe size $|U|$\\
$d$ & maximal set size with guaranteed listing\\
$m$ & filter length\\
$k$ & number of hash functions \\
$M$ & binary matrix representing the hash functions\\
\hline
\end{tabular}
\caption{Summary of main notations}
\label{table:notations}
\end{table}

\section{Designing IBLTs with listing guarantees}\label{sec: Designing IBLT}
In the current and the following sections we present constructions for IBLTs with listing guarantees. Here we present constructions based on three known methodologies such as Bloom filters, linear codes, and combinatorial structures.  Later, in \autoref{section_recursive_constructions} we describe new recursive constructions we develop.   
A detailed summary of the constructions and their relevant parameters is given in \autoref{table:results}. 

\subsection{Bloom Filters with FPFZ}
\label{section_bloom_constructions}
Here we define the mapping matrix required by the Bloom filter with a FPFZ (false positive free zone) and show how to derive a \decodable{} matrix from it.

We start by articulating the set of mapping matrices that defines the family of Bloom filters with $\del{n,d}$-FPFZ.
Let $S \subseteq U$ be a set with at most $d$ elements.
A Bloom filter with the set $S$ does not have false positives if any other element $u \in U \setminus S$ has an index $i \in \sbr{1, m}$ such that $u_i=1$ and $s_i=0$ for all $s \in S$.
It can be shown that the this requirement is captured by the following definition.

\begin{definition}[$d$-FPF matrix]\label{def: d-FPF}
An $m\times n$ binary matrix $M$ is called $d$-false-positive-free ($d$-FPF), if any $m \times \del{d+1}$ sub-matrix of $M$ contains all the possible $d+1$ unit vectors of length $d+1$ within its rows.
\end{definition}
Since all the unit vectors of length $d+1$ are contained within the rows of any $m\times \del{d+1}$ sub-matrix of a $d$-FPF matrix, any such $m\times \del{d+1}$ sub-matrix contains the identity matrix $I_{d+1}$ (up to row permutation) as its sub-matrix. Note that from a coding theory perspective, a $d$-FPF matrix forms a \emph{zero-false-drop} code of order $d$ ($ZFD_d$). This family of codes was studied for several applications in files retrieval, data communication, group testing and magnetic memories~\cite{superimposedcodes, eppstein2007improved}. 
This observation simply verifies the following theorem.

\begin{theorem}\label{th:FPFZ to LFFZ}
Let $M$ be an $m \times n$ matrix. If $M$ is a $d$-FPF matrix, then $M$ is also a \decodable[(d+1)] matrix.
\end{theorem}

We mark that the set of $d$-FPF matrices is not equivalent to the set of \decodable[(d+1)] matrices.
The following example demonstrates that there exist \decodable[(d+1)] matrices which are not $d$-FPF matrices.

\begin{figure}[h!]
\centering
\begin{minipage}{.78\textwidth}
\begin{example}
The matrix $M$ from \autoref{example:matrix} is \decodable[3].
To show that $M$ is not $2$-FPF, we present a $5 \times 3$ sub-matrix that does not contain all the possible unit vectors of length 3.
Consider the sub-matrix for $S'=\cbr{1, 2, 5}$, presented on the right.
This matrix does not contain the unit vector $\del{0, 1, 0}$, and hence $M$ is not $2$-FPF.
From a Bloom filter perspective, if the filter contains the set $\cbr{1, 5}$, querying for the element 2 results in a false positive, as both the first and fourth bits that element 2 is mapped to equal 1.
\end{example}
\end{minipage}\hspace{.5cm}%\hfill
\begin{minipage}{.12\textwidth}
\vspace{-.5cm}
    \[
    \begin{bmatrix}
    1 & 1 & 0 \\
    0 & 0 & 1 \\
    1 & 0 & 0 \\
    0 & 1 & 1 \\
    0 & 0 & 0 \\
    \end{bmatrix}
    \]
\end{minipage}
\end{figure}

According to \autoref{th:FPFZ to LFFZ} upper bounds on $\minm{n}{d}$ can be derived from constructions of Bloom filters with FPFZ. For a Bloom filter with a FPFZ, \cite{EGH} suggests a solution called EGH Bloom filter with \(m=O\del{d^2\cdot \log_2{n}}\) such that $k$ is the minimal number of the first consecutive primes $q_1 = 2, q_2 = 3, \ldots, q_k$ that their product satisfies $\prod_{i=1}^{k} q_i \ge n^d$. This implies that \(k=O\del{d\cdot\log_2{n}}\).
A solution based on Orthogonal Latin Squares (OLS) was suggested in~\cite{FPFZ} with \(m=\del{d+1}\sqrt{n}\)  and \(k=d+1\).
The authors presented also a polynomial based filter (POL) of a similar result with a tunable parameter that can trade $k$ for memory $m$. The next corollary summarizes these claims.
%we have the following corollaries.

\begin{corollary}\label{cor: LFFZ from FPFZ}
For any integers $n, d$, the followings hold,
\begin{enumerate}
    \item $\minmk{n}{d}{\min \{k | \prod_{i=1}^{k} q_i \ge n^d\}} \le O\del{\del{d-1}^2 \cdot \log_2{n}}$ where $q_1 = 2, q_2 = 3, \ldots, q_k$ refer to the $k$ smallest primes.
    \item $\minmk{n}{d}{d} \le d \sqrt{n}$.
\end{enumerate}
\end{corollary}

\subsection{Constructions based on Linear Codes}
\label{section_linear_codes_constructions}
In \autoref{sec: problem definition} we defined stopping sets and the stopping distance of a matrix (\autoref{def: stopping set}). The stopping distance was studied in the context of parity-check matrices of linear codes~\cite{stoppingSets, schwartz2006ss, etzion2006stopping}.  We show how to utilize results from previous works on stopping sets of parity-check matrices of known error-correcting codes.

%Let $F_q$ be an alphabet of size $q$. A code is a nonempty subset $\mathcal{C}$ of $F_q^n$ with $M$ elements. The elements of the code are called codewords. For a code $\mathcal{C}$, the minimum distance of $\mathcal{C}$, denoted by $d$, is the minimum Hamming distance between any two distinct codewords of $\mathcal{C}$. Where the Hamming distance between two words of the same length is defined to be the number of coordinates in which these words differ. A code $\mathcal{C}$ with $M$ codewords of length $n$ and minimum distance $d$ is called linear if it is a linear subspace of $F_q^n$. The dimension of a liner code $\mathcal{C}$ is the dimension of $\mathcal{C}$ as a liner subspace of $F_q^n$ over $F_q$. A linear code $\Code \subseteq F_q^n$ with codewords of length $n$, dimension $k$ and minimum distance $d$ is denoted by $[n,k,d]_q$.

A length-$n$ binary code is a nonempty subset $\mathcal{C}\subseteq \{0,1\}^n$ with $M$ elements. The elements of the code are called \emph{codewords}. For a code $\mathcal{C}$, the minimum Hamming distance of $\mathcal{C}$, denoted by $d(\mathcal{C})$, is the minimum Hamming distance between any two distinct codewords of $\mathcal{C}$, where the Hamming distance between two words of the same length is defined to be the number of coordinates in which these words differ. A code $\mathcal{C}$ with $M$ codewords of length $n$ and minimum Hamming distance $d$ is called linear if it is a linear subspace of $\{0,1\}^n$. The dimension of a linear code $\mathcal{C}$ is the dimension of $\mathcal{C}$ as a linear subspace of $\{0,1\}^n$ over $\{0,1\}$. A linear code $\Code \subseteq \{0,1\}^n$ with codewords of length $n$, dimension $k$, and minimum Hamming distance $d$ is denoted by $[n,k,d]$. The \emph{redundancy} of the code is defined to be $n-k$.

% A parity-check matrix of a liner $[n,k,d]_q$ code $\mathcal{C}$ is an $r\times n$ matrix $H$ such that for any $x\in F_q^n$, $x$ is a codeword of $\mathcal{C}$ if and only if $Hc^T = 0$. Note that $H$ is not unique and if $H$ is of full rank, then $r=n-k$ and otherwise $r>n-k$.

A parity-check matrix of a binary linear $[n,k,d]$ code $\mathcal{C}$ is an $r\times n$ binary matrix $H$ such that for any ${\bf c}\in \{0,1\}^n$, ${\bf c}$ is a codeword of $\mathcal{C}$ if and only if $H{\bf c}^T = 0$. It is well known that every linear code can be represented by a parity-check matrix $H$, and note that $H$ is not unique. In case $H$ is of full rank, then $r=n-k$ and otherwise $r>n-k$.

% \begin{example}\label{ex: Extended Hamming}
% The binary Extended Hamming code $\mathcal{H}^*(8)$ of length $n=8$ is a linear $[n=8,k=4,d=4]$ code with the following parity-check matrix,
% \[
% H = 
% \begin{bmatrix}
% 0 & 1 & 1 & 1 & 1 & 0 & 0 & 0\\
% 1 & 0 & 1 & 1 & 0 & 1 & 0 & 0\\
% 1 & 1 & 0 & 1 & 0 & 0 & 1 & 0\\
% 1 & 1 & 1 & 0 & 0 & 0 & 0 & 1\\
% \end{bmatrix}
% \]
% and redundancy $r=n-k=4$. 
% \end{example}

For a linear code $\Code$, the \emph{stopping redundancy} $\rho\del{\Code}$ is defined as the minimal number of rows in a parity-check matrix $H$ for $\Code$ such that $d\del{\Code}=s\del{H}$.
Studying the stopping distance of parity check matrices and the stopping redundancy of codes was motivated by the observation that the performance of linear codes under iterative decoding over the binary erasure channel is determined by the size of the smallest stopping set in the Tanner graph, i.e., of the parity check matrix~\cite{schwartz2006ss}.
In 
Based on this discussion we derive the following result, proved in~\autoref{app:A}.
%We now show how a \decodable{} matrix can be constructed from linear codes, and then generalize to $q$-ary linear codes.
%We show next how \decodable{} matrices can be constructed from linear codes.

\begin{theorem}\label{th:linear codes to d-decodable}
For any binary linear $\sbr{n, k, d}$ code $\Code$ with stopping redundancy $\rho=\rho\del{\Code}$ and minimum distance $d>2$, there exists a \decodable[(d-1)] matrix of size $\rho \times n$.
\end{theorem}
% \begin{example}
% It was proven in~\cite{etzion2006stopping} that the stopping redundancy of the $[8,4,4]$ code $\mathcal{H}^*(8)$ (see \autoref{ex: Extended Hamming}) is $\rho(\mathcal{H}^*(8)) = 5$ and the corresponding parity-check matrix was given in~\cite{schwartz2006ss} and equals
% \[
% H' = \begin{bmatrix}
% 1 & 0 & 1 & 0 & 1 & 0 & 1 & 0\\
% 0 & 1 & 0 & 1 & 0 & 1 & 0 & 1\\
% 0 & 0 & 1 & 1 & 0 & 0 & 1 & 1\\
% 0 & 0 & 0 & 0 & 1 & 1 & 1 & 1\\
% 1 & 1 & 1 & 1 & 0 & 0 & 0 & 0
% \end{bmatrix}
% \]
% It can be verified that $Hc^T=0$ if and only if $H'c^T=0$ and that the smallest stopping set in $H'$ is of size $4$ which implies that $H'$ is a \decodable[3] matrix with $\rho(\mathcal{H}^*(8)) = 5$ rows. 
% \end{example}

\autoref{th:linear codes to d-decodable} implies that $d$-decodable matrices exist for an infinite number of parameters. However, in order to find a $d$-decodable matrix with the smallest possible number of rows, we need to find an $[n,k,d+1]$ code $\Code$ with minimum stopping redundancy $\rho\del{\Code}$. Even though the stopping redundancy was studied extensively for a variety of families of linear codes, the problem of finding the code $\Code$ with minimum stopping redundancy $\rho\del{\Code}$ is still open and has not been addressed for most parameters.
In fact, one should note that the problems of stopping redundancy for parity check matrices of error-correcting codes and $d$-decodable matrices for IBLT with LFFZ are not equivalent. The goal in both of these problems is to find matrices with a prescribed value for their stopping distance, while minimizing the number of rows in the matrix. However, while in parity check matrices of error-correcting codes, minimizing the number of rows in the matrix is only a secondary goal to the main one of minimizing the rank of the matrix, for $d$-decodable matrices minimizing the number of rows is the only goal. In fact, for this reason the stopping redundancy has been studied for specific families of codes and not as a goal to minimize the stopping redundancy. More than that, we may also consider square invertible matrices as $d$-decodable matrices (see~\autoref{th:basic}(3)), which, as parity check matrices, result with trivial codes. On the other hand, if the stopping redundancy of an error-correcting code is greater than $n$, such as Reed-Muller codes~\cite{etzion2006stopping} and the Golay codes~\cite{schwartz2006ss}, then it is of no use in our problem. 

In~\autoref{cor:from linear}, we conclude two general upper bounds based upon results on the stopping redundancy of the Hamming and Simplex codes. We did not list other general results since their stopping redundancy value was larger than the code length. Furthermore, although extensive research has been done in the area of stopping redundancy of linear codes, there is only a few works that consider binary codes and yield interesting results under our framework of \decodable[d] matrices. Several works studied the minimum distance of several families of low-density parity-check (LDPC) matrices; see e.g.~\cite{Esmaeili_LDPC,rosnes2014minimum,laendner2007ldpc} and references therein. The authors of~\cite{Esmaeili09,Esmaeili_LDPC} studied proper and improper array codes and cases in which the stopping distance and minimum distance of these codes are and are not the same. In~\cite{rosnes2014minimum}, the authors studied the stopping distance of array LDPC matrices and a table with numeric results for several small values of $m,n,$ and $k$, that were found using a computer search, is given. The stopping distance of parity-check matrices for some families of LDPC codes was also studied in~\cite{laendner2007ldpc} and a construction for \decodable[(5,3)] matrices is presented.
Recently, the case $d=\Theta\del{n}$ and the family of matrices that contains the identity matrix as a sub-matrix were studied in~\cite{linial2022bounds}.
According to \autoref{th:linear codes to d-decodable}, more results which consider the stopping redundancy of infinite families of binary linear codes provide upper bound on $\minm{n}{d}$ and are summarized in the next corollary.
%In the next corollary we summarize the results on $m^*$ that can be derived from the stopping redundancy of binary linear codes using \autoref{th:linear codes to d-decodable}.  

%An $m q \times q^2$ matrix $H\del{q,m}$ for $\del{m,q}$-regular LDPC code (i.e. $k=q$ in our terms).A table with stopping distance results for $7 \leq q \leq 79$ and $4 \leq m \leq 7$ is given, found by computer search.It seems like increasing $q$ over some lower value does not improve the stopping distance, but for $m$ it does.Stopping distances varies between 8-24, for large enough values of $q$ we can derive \decodable[16]. \cite{rosnes2014minimum}

%In the next corollary we summarize the results on $m^*$ that can be derived from the stopping redundancy of binary linear codes using \autoref{th:linear codes to d-decodable}. It should be noted that although extensive research has been done in the area of stopping redundancy, the following corollary considers only the works that yields an interesting result under our framework of \decodable[d] matrices.   
\begin{corollary}\label{cor:from linear}
The following claims hold.
\begin{enumerate}
%    \item If there exists an $[n,k,3]$ code, then $\minm{n}{d=3}\le 2(n-k)-3$.
%    \item $\minm{n}{d=3} \leq 2\log n -1$~\cite{}.
    \item $\minm{n}{d=3} \leq 2\lceil\log_2 n\rceil-1$~\cite{etzion2006stopping,schwartz2006ss}.
    \item $\minm{n=2^\ell-1}{d=2^{\ell-1}-1} \le 2^\ell-\ell -1 \approx n -\log_2 n$, for $\ell\geq 1$~\cite{etzion2006stopping}.
    \item For all odd prime $q$, we have $m^*(n=q^2,d=3,k=2) \le 2q$, 
    {$m^*(n\hspace{-0.25ex}=\hspace{-0.25ex}q^2,d\hspace{-0.25ex}=\hspace{-0.25ex}5,k\hspace{-0.25ex}=\hspace{-0.25ex}3)\hspace{-0.5ex}\le\hspace{-0.5ex} 3q,m^*(n\hspace{-0.25ex}=\hspace{-0.25ex}q^2,d\hspace{-0.25ex}=\hspace{-0.25ex}7,k\hspace{-0.25ex}=\hspace{-0.25ex}4) \hspace{-0.5ex}\le\hspace{-0.5ex} 4q$}~\cite{Esmaeili09}.
    \item For all $v\hspace{-0.25ex}\geq\hspace{-0.25ex} 1$, $m^*(n\hspace{-0.25ex}=\hspace{-0.25ex}(2v+1)(3v+1),d\hspace{-0.25ex}=\hspace{-0.25ex}5,k\hspace{-0.25ex}=\hspace{-0.25ex}3) \hspace{-0.25ex}\le\hspace{-0.25ex} 6v\hspace{-0.25ex}+\hspace{-0.25ex}3$~\cite{laendner2007ldpc}.

\end{enumerate}
\end{corollary}

Yet another connection using the stopping redundancy of BCH codes with minimum distance 5 is shown in the next corollary. The proof is given for completeness as we could not find it elsewhere. 
\begin{corollary}\label{cor: from BCH}
It holds that $\minm{n}{d=4} \leq \minmk{n}{d=4}{k=2\ceil{\log(n+1)}} \le 4\ceil{\log(n+1)}$.
\end{corollary}
\begin{proof}
Let $H$ be an $(n-k)\times n$ parity check matrix of an $[n,k,5]$ linear code $\Code$. Consider the $2(n-k)\times n$ matrix $\widehat{H} = [H:\overline{H}]$, where $\overline{H}$ is the complement matrix of $H$ and the matrices $H$ and $\overline{H}$ are concatenated by the columns. Since the minimum distance of $\Code$ is 5, every 4 columns in the matrix $H$ are linearly independent and in particular there exists a row of weight 1 or 3. This directly implies that these 4 columns in the matrix $\widehat{H}$ have a row of weight 1. Lastly, an explicit construction can be given by a parity check matrix of a BCH code with minimum distance 5~\cite{roth_2006}.
\end{proof}

\subsection{Steiner Systems}\label{section_steiner_constructions} %
In this section we derive results on \decodable[d] matrices based upon results from Steiner systems. This family of a combinatorial design is first defined. 
\begin{definition} Let $Q$ be an $m$-set (points) and let $B$ be a collection of $k$-subsets (blocks) of $Q$. The pair $(Q,B)$ is called a \textbf{Steiner system} $S(t,k,m)$ if any $t$-subset of $Q$ is contained in exactly one block of B.
\end{definition}
For a Steiner system  $S=S(t,k,m)$, we denote by $n$ the number of blocks in $S$ (i.e. $n\triangleq |B|$). The \emph{incidence matrix} of a Steiner system $S=S(t,k,m)$ is a binary matrix $M$ of size $m\times n$, namely its number of rows equals the number of points and its number of columns equals the number of blocks in the system.  
For $1\le i\le m$ and $1\le j \le n$, $M(i,j)=1$ if and only if the $i$-th point of $Q$ is contained in the $j$-th block of $B$.

Steiner systems were also studied in the context of stopping sets and stopping redundancy of linear codes and the following theorem was proven in~\cite{kashyap2003stopping}.
\begin{theorem}\label{the: stopping set in an incidence structre}
Let $M$ be an incidence structure of points and blocks such that each block contains exactly $k$ points, and each pair of distinct blocks intersect in at most $\gamma$ points. If $S$ is a stopping set in $S$, then $|S| \ge k / \gamma + 1$.
\end{theorem}

% \begin{example}
% Consider the incidence matrix $M$ of the $(2,3,7)$ Steiner system that was given in \autoref{examp: Fano}. 
% It can be readily verified that $M$ is \decodable[3]. In fact, in this case the bound is also tight since the last four columns of $M$ form a stopping set of size four.   
% \end{example}

\autoref{the: stopping set in an incidence structre} implies that the minimum size of any stopping set in a Steiner system $S=S(t,k,m)$ is at least $\frac{k}{t-1}+1$. Hence, in order to increase the stopping distance of the incidence matrix of the design, it is preferable to use a smaller value of $t$, and in particular we use the existing constructions for $t\leq 3$. There are several families of Steiner systems that have been studied for this special case and some of the relevant results for our problem, which improve upon \autoref{th:basic}(4), are summarized in the next corollary. 

\begin{corollary}\label{cor: Steiner and IBLT}
Let $S=S(t,k,m)$ be a Steiner system. The incidence matrix of $S$ is a \decodable[\del{\left\lceil\frac{k}{t-1}\right\rceil}] matrix. In particular, we have that
\begin{enumerate}
    \item $m^*(n,d = 3, k=3) \le \left\lceil\frac{1+\sqrt{1+24n}}{2}\right\rceil + 3 =\Theta(\sqrt{n}).$
    \item $ m^*\left(n,d=\left\lceil\frac{q+1}{2}\right\rceil,k=q+1 \right)= O(d^2\sqrt[3]{n}).$
    \end{enumerate} 
\end{corollary}

The results of~\autoref{cor: Steiner and IBLT} follow from \autoref{the: stopping set in an incidence structre} and the following two families of Steiner systems; (1) A Steiner system $S=S(2,3,m)$, $m(\bmod{6})\in \{1,3\}$, also called a \emph{Steiner triple system of order $m$}~\cite{colbourn1999triple}, and (2) Steiner system $S=S(3,q+1,q^\alpha +1)$ for prime power q and any integer $\alpha\ge 2$~\cite{blanchard1995construction}. The full details appears in~\autoref{app:A}.

\subsection{Covering Arrays}\label{section_covering_arrays_constructions}
So for we discussed two families of matrices. While the first one, used for the construction of IBLTs, requires every set of $d$ columns to contain at least one row of weight one, the second one, used for Bloom filters, imposes a stronger property for the $d$ columns to contain all $d$ rows of weight one. Yet another family of matrices requires even a stronger property in which every $d$ columns contain \emph{all} possible $2^d$ rows. While we are interested in the first family with the weakest property, apparently the results for the last family of matrices with the strongest property lead to interesting and non-trivial results for \decodable[d] matrices. Let us first define mathematically this family of matrices. 
\begin{definition}
A binary $m\times n$ matrix $M$ is called a binary covering array with strength $d$ if any $m\times d$ sub-matrix contains all the possible $2^d$ binary vectors of length $d$ within its rows (at least once).   
\end{definition}
%The next connection is simple.
\begin{lemma}\label{lem: covering arrays and IBLT codes}
Any binary covering array of strength $d$ is a \decodable[d] matrix.
\end{lemma}
A comprehensive survey of binary covering arrays can be found in~\cite{CoveringArrays}. We give here one of the results, which is relevant to our discussion and provides another upper bound on $m^*(n,d)$. 
\begin{theorem}\label{the: upper bound covering arrays}
There exists a binary $m\times n$ covering array of strength $d$ such that
$$m^*(n,d) \leq m\le \frac{d}{\log_2\left(\frac{2^d}{2^d-1}\right)}\log_2(n).$$
\end{theorem}

\section{Recursive constructions and upper bounds}\label{section_recursive_constructions}
In this section we present recursive upper bounds on $m^*(n,d)$ and $m^*(n,d,k)$ for any $n,d$ and $k$. 
Then, using these bounds, we design a family of recursive constructions of \decodable[d] and \decodable[(d,k)] matrices with $n$ columns, which are suitable for any $n,d,$ and $k$. In addition, we derive closed form upper bounds on $m^*(n,d)$ by selecting specific matrices as the seed of the recursive constructions and analyze the number of rows in the \decodable[d] matrices obtained by the  constructions. It is shown that for $d=3$, our construction achieves \decodable[3] matrices with $n$ columns and $m=\frac{3}{\log_2 3}\lceil\log_2 (n)\rceil$ rows. Recall that the minimum number of rows in the \decodable[3] matrices we presented so far was $m=2\lceil\log_2 n\rceil-1$. Hence, our recursive construction outperforms any other construction for \decodable[3] matrices known to us, and thus leads to a better upper bound for $m^*(n,3)$. 

For a matrix $M$ let $\Rows{M}$, $\Cols{M}$ be the number of rows, columns in $M$, respectively.
We start by presenting a basic recursive upper bound on the minimum number of rows in a \decodable[d] matrix. % with $n$ columns. 

\begin{theorem}\label{th: recursive improved}
For integers $d\ge 3$ and $n\geq 1.5(d+1)$, it holds that
$$
m^*(n,d)\le \min_{2\le i \le \lceil {n}/2\rceil}\left\{ m^*\left(\lceil {{n}}/{i}\rceil , d \right) + i\cdot  m^*\left(\lceil {n}/i\rceil, \lfloor d/2\rfloor \right)\right\}.
$$
\end{theorem}

\begin{proof}
We prove the claim by showing that
$m^*(n,d)\le m^*\left(\lceil {{n}}/{i}\rceil , d \right) + i\cdot  m^*\left(\lceil {n}/i\rceil, \lfloor d/2\rfloor \right)$ for any ${2\le i \le \lceil {n}/2\rceil}$.
We present the proof only for $i=2$, while the generalization for the other values of $i$ can be readily verified.

For any $n,d$ let $H^n(d)$ be an optimal \decodable[d] matrix with $n$ columns and minimum number of rows, that is, by definition, $\Rows{H^n(d)}=m^*(n,d)$. We prove the theorem by constructing a \decodable[d] matrix $H$ with $n$ columns and  $m^*\left(\lceil n/2\rceil , d \right) + 2m^*\left(\lceil n/2\rceil, \lfloor d/2\rfloor \right)$ rows, and then conclude 
$$
\hspace{-0.3ex}m^*(n,d) \hspace{-0.3ex}=\hspace{-0.3ex} \Rows{H^n(d)}\hspace{-0.3ex}\le\hspace{-0.3ex} \Rows{H} \hspace{-0.3ex}=\hspace{-0.3ex} m^*\left(\left\lceil \frac{n}{2}\right\rceil , d \right) + 2 m^*\left(\left\lceil \frac{n}{2}\right\rceil, \left\lfloor \frac{d}{2}\right\rfloor \right)\hspace{-0.3ex}.
$$

Let 
\[
H = 
\begin{bmatrix}
 H^{\lceil n/2\rceil }(\lfloor d/2 \rfloor)      &  \textbf{0}       \\
\textbf{0}      &   H^{\lceil n/2\rceil}(\lfloor d/2 \rfloor)       \\
H^{\lceil n/2\rceil}(d)         &  H^{\lceil n/2\rceil}(d)          \\
\end{bmatrix},
\]
where \textbf{0} is the zero matrix of size ${\Rows{H^{\lceil n/2\rceil }(\lfloor d/2 \rfloor)}\times \lceil n/2\rceil}$. Clearly, $\Rows{H} = m^*\left(\lceil n/2\rceil , d \right) + 2 m^*\left(\lceil n/2\rceil, \lfloor d/2\rfloor \right)$. In addition, if $n$ is even then $\Cols{H}=n$ and otherwise $\Cols{H}=n+1$. \ Note that if $H$ is \decodable[d] then any matrix $H'$ that is obtained from $H$ by erasing one column is also \decodable[d] and hence it is sufficient to prove that for even integer $n$, the matrix $H$ is \decodable[d].

It is easy to verify that all the columns of $H$ are unique and hence it is sufficient to prove that any set of at most $d$ columns of $H$ contains a row of weight one. We present the proof for exactly $d$ columns while the same proof holds for a smaller number of columns. Any $d$ columns taken altogether either from the first or the last $n/2$ columns of $H$ must have row of weight one in the sub-matrix $H^{\lceil n/2\rceil}(d)$ since this sub-matrix is \decodable[d].
Otherwise, w.l.o.g., the number of columns taken from the first $n/2$ columns of $H$ is at most $\lfloor d/2\rfloor$. Since the sub-matrix $H^{\lceil n/2\rceil }(\lfloor d/2 \rfloor)$ is \decodable[\lfloor d/2 \rfloor], the first $n/2$ columns have a row of weight one in one of the first $\Cols{H^{\lceil n/2\rceil}(d)}$ rows. The columns from the last $n/2$ columns have zeros in all the first $\Cols{H^{\lceil n/2\rceil}(d)}$ entries and thus the $d$ columns contain a row of weight one. 
\end{proof}

Using the ideas from the proof of \autoref{th: recursive improved} we present a recursive construction of a \decodable[d] matrix with $n$ columns.
This construction is based on the existence of a \decodable[\lfloor d/2\rfloor] and \decodable[d] matrices with $\lceil n/i\rceil$ columns.
Recall that for any $n\ge 1$, the all-one row vector of length $n$ is \decodable[1] and any matrix with $n$ unique non-zero columns is \decodable[2].

\begin{cnstr}\label{const: improved d-decodable}
Let $d\ge 3, n\ge 1$. Let $M^n(d)$ be a matrix with $n$ columns  defined recursively as follows. For any $n\le d$, $M^n(d) = I_n$, the identity matrix of size $n$, and for all $d<n<1.5(d+1)$,
\[
M^{n}(d) = 
\begin{bmatrix}
I_{n-1}  &  \textbf{1} \\
\end{bmatrix}, 
%\;\;\;
%M^{2d+1}(d) = 
%\begin{bmatrix}
%\textbf{0}  &  M^{d+1}(\lfloor d/2 \rfloor) \\
%I_d         &  M^{d+1}(d)        \\
%\end{bmatrix},
\]
where $\textbf{1}$ is the all-one column vector.
For any $n\ge 1.5(d+1)$,  define
\[
M(n,d,i) = 
\begin{bmatrix}
\begin{matrix}
 M^{\lceil n/i\rceil }(\lfloor d/2 \rfloor)  & \textbf{0}  &  \textbf{0} \\
\textbf{0} & \ddots      &  \textbf{0} \\
\textbf{0} & \textbf{0}  &   M^{\lceil n/i\rceil }(\lfloor d/2 \rfloor)  \\
 M^{\lceil n/i\rceil }( d )    & \dots       &   M^{\lceil n/i\rceil }( d )    \\
\end{matrix}
\end{bmatrix}
\]
where \textbf{0} is the zero matrix of size ${\Rows{M^{\lceil n/i\rceil }(\lfloor d/2 \rfloor)}\times \lceil n/i\rceil}$. The matrix $M^n(d)$ is defined by selecting the value of $i$ for which $M(n,d,i)$ has the minimum number of rows and then erasing columns to obtain a matrix with exactly $n$ columns. 
\end{cnstr}

By modifying \autoref{const: improved d-decodable} such that we always set $i=2$, the next upper bound for $m^*(n,d)$ can be derived.
For proof details see \autoref{app:A}.

\begin{theorem}
\label{theorem_construction_A}
For any $n\ge 1, d\ge 3$, we have that $m^*(n,d)= O\del{\log_2^{\floor{\log_2 d}} n}$,
and in particular, $m^*(n,3)\le 2\ceil{\log_2 n} -1$.
\end{theorem}

In the rest of this section, we develop a closed form expression for the number of rows in a \decodable[3] matrix with $n$ columns obtained by the improved construction. This construction outperforms any other construction for \decodable[3] matrices known to us and achieves a \decodable[3] matrix with $n$ columns and $m = \frac{3}{\log_2 3}\lceil \log_2 (n)\rceil$ rows.

%with respect to minimizing the number of rows and the number of rows in the obtained \decodable[3] matrix with $n$ columns is $m = \frac{3}{\log 3}\log (n)$.

%which will yield a better upper bound on $m^*(n,3)$.
%Next, we present an improvement for \autoref{const: basic d-decodable} which will yield a better upper bound on $m^*(n,3)$.

% \begin{theorem}\label{th: improved d-decodable}
% For any $n\ge 1$, the matrix $M^n(d)$ obtained by \autoref{const: improved d-decodable} is \decodable[d]. In addition, for $n\ge 1.5(d+1)$,
% $$
% \Rows{M^n(d)} = \min_{2\le i \le \ceil{\frac{n}{2}}}\cbr{ \Rows{M^{\ceil{\frac{n}{i}} }(d)} + i\cdot  \Rows{M^{\ceil{\frac{n}{i}} }\del{\floor{\frac{d}{2}}}}}
% .$$ 
% \end{theorem}
% The correctness of \autoref{const: improved d-decodable} is based on the proof of \autoref{th: improved d-decodable} and hence it is omitted. 

In case $d=3$, for any $i$ the sub-matrix $ M^{\lceil n/i\rceil }(\lfloor d/2 \rfloor)$ in \autoref{const: improved d-decodable} is the all-one row vector of length $\lceil n/i\rceil$. Using this observation, \autoref{const: improved d-decodable} can be modified such that the recursion step will be a function of the number of rows instead of the number of columns. In the following construction of \decodable[3] matrices, we further improve \autoref{const: improved d-decodable}  using this observation and by increasing the number of columns with additional three columns for any given number of rows.   
\begin{cnstr}\label{const: improved 3-decodable}
Let $M_m$ for $m\ge 3$ be a matrix with $m$ rows which is defined recursively as follows.
\[
M_3 = \begin{bmatrix}
I_3  &  \textbf{1} \\
\end{bmatrix}
\;\;\;
M_4 = \begin{bmatrix}
\textbf{0}  &  \textbf{1} \\
I_3         &  M_3        \\
\end{bmatrix}
\]  and for $m\ge5$, 
\[
M_m = 
\begin{bmatrix}
\begin{matrix}
\textbf{0}  \\
\vdots      \\
\textbf{0}  \\
I_3
\end{matrix}
& \vline &
\begin{matrix}
\textbf{1} & \textbf{0}  &  \textbf{0} \\
\textbf{0} & \ddots      &  \textbf{0} \\
\textbf{0} & \textbf{0}  &  \textbf{1} \\
M_{m-i}   & \dots       &  M_{m-i}   \\
\end{matrix}
\end{bmatrix}
\]
where $i$ is chosen such that the number of columns is maximized, i.e., 
\[
i = \argmax_{i \in \sbr{2,m-3}}{\del{i\cdot \Cols{M_{m-i}}}}.
\]
\end{cnstr}
%Note that in~\autoref{const: basic 3-decodable} we simply fixed the value of $i$ to be $i=2$ for any $m\ge 5$.
\begin{lemma}\label{th: const_C_th}
For any $m\ge 3$, the matrix $M_m$ obtained by \autoref{const: improved 3-decodable} is \decodable[3]. 
\end{lemma}
\begin{proof}
It can be readily verified that $M_3$ and $M_4$ are \decodable[3] matrices. For $m\ge 5$, all the columns of $M_m$ are unique and hence it is sufficient to show that any set of $3$ columns contains a row of weight one. Denote the indices of these columns by $j_1<j_2<j_3$. Let $\alpha$ denote the number of columns that are taken from the three left-most columns of $M_m$ and consider the following  distinct cases.
\begin{enumerate}
    \item If $\alpha = 3$ then they clearly contain a row of weight one.
    \item If $\alpha = 2$ then the $j_3$-th column contains an entry with $1$ in the first $i$ rows. Since the first $i$ entries of the other two columns equal to zero, this implies a row of wight one.
    \item  If $\alpha=1$, the $j_2$-th and the $j_3$-th columns belong to either a single sub-matrix $M_{m-i}$ or to two distinct sub-matrices. It is possible to see that if the two columns belong to distinct sub-matrices, the three columns have a row of weight one. Now assume the two columns belong to the same sub-matrix. Here we note that the first three columns of the sub-matrix $M_{m-i}$ are identical to the first three columns of $M_m$ while ignoring the first $i$ rows of the matrix. If the $j_1$-th column is identical to the $j_2$-th, $j_3$-th column (while ignoring the first $i$ rows) then there exists a row for which only the $j_3$-th, $j_2$-th column contains a 1 entry, respectively. Otherwise, the $j_1$-th column is identical to neither the $j_2$-th nor the $j_3$-th column (while ignoring the first $i$ rows), but there exists another column in this sub-matrix which is identical to this column. Then, these columns cannot form a stopping set since otherwise there will be a stopping set in the sub-matrix $M_{m-i}$. 
    \item If $\alpha=0$, the existence of a row with weight one follows from the correctness of \autoref{const: improved d-decodable}.
\end{enumerate}
\end{proof}

Before evaluating the relation between $m$ and $\Cols{M_m}$, we show that the integer $i$ from~\autoref{const: improved 3-decodable} equals to $3$ for any $m\ge9$.

\begin{lemma}\label{lem: i=3}
For $m \geq 9$, $\argmax_{i \in \sbr{2,m-3}}{\del{i \Cols{M_{m-i}}}} = 3.$
\end{lemma}

\autoref{lem: i=3} is proved in~\autoref{app:A}.
Using~\autoref{lem: i=3} we can analyze the number of columns in $M_m$ and obtain the following upper bound on $m^*(n,3)$.
The proof can be found in~\autoref{app:A}.

\begin{theorem}\label{th:d3construction}
For any $n \ge 3$ we have that
$\minm{n}{3} \le \recresult$.
\end{theorem}

The bound of \autoref{th: recursive improved} and \autoref{const: improved d-decodable} can be easily modified in order to construct \decodable[(d,k)] matrices by replacing the \decodable[d] and the \decodable[\lfloor d/2 \rfloor] matrices with $\lceil n/i\rceil$ columns that are used by the construction with a \decodable[(d,k_1)] matrix and a \decodable[(\lfloor d/2 \rfloor, k_2)] matrix such that $k_1+k_2 = k$.
The modified bound and~\autoref{const: basic (d,k)-decodable} are given next.

\begin{theorem}\label{th: recursive d,k improved}
For positive integers $d,k$ and $n> d$, it holds that
$$
m^*(n,d,k)\le \hspace{-0.3ex}\min_{\substack{(i,k_1,k_2):\\ 2\le i \le \lceil \frac{n}{2}\rceil\\ k_1+k_2=k}} \hspace{-0.2ex}\left\{ m^*\left(\left\lceil \frac{n}{i}\right\rceil , d, k_2 \right) + im^*\left(\left\lceil \frac{n}{i}\right\rceil, \left \lfloor\frac{d}{2}\right\rfloor,k_1 \right) \right\}.
$$
\end{theorem}

\begin{cnstr}\label{const: basic (d,k)-decodable}
For $n,k\ge 1$ and $d\ge 3$ let $M^n(d,k)$ be a matrix with $n$ columns which is defined recursively as follows. For any $n\le d$, the matrix $M^n(d)$ is given by % = I_n$, the identity matrix of size $n$, 
\[
M^{n}(d) = 
\begin{bmatrix}
I_d   \\
\textbf{1}_{k-1\times d}
\end{bmatrix}, 
\]
where $\textbf{1}_{k-1\times d}$ is the all-one matrix of size $k-1\times d$. For $n\ge d+1$ and $k$, let $i,k_1,k_2$ be such that $k_1+k_2=k$ and $2\le i\le \lceil n/2\rceil$. We define
\[
M(n,d,i,k_1,k_2) = 
\begin{bmatrix}
\begin{matrix}
 M^{\lceil n/i\rceil }(\lfloor d/2\rfloor,k_1)  & \textbf{0}  &  \textbf{0} \\
\textbf{0} & \ddots      &  \textbf{0} \\
\textbf{0} & \textbf{0}  &   M^{\lceil n/i\rceil }(\lfloor d/2 \rfloor,k_1)  \\
 M^{\lceil n/i\rceil }( d,k_2 )    & \cdots       &   M^{\lceil n/i\rceil }( d ,k_2)    \\
\end{matrix}
\end{bmatrix}
\]
where \textbf{0} is the zero matrix of size ${\Rows{M^{\lceil n/i\rceil }(\lfloor d/2 \rfloor),k_1}\times \lceil n/i\rceil}$. The matrix $M^n(d,k)$ is defined by selecting the values of $i,k_1$, and $k_2$ for which $M(n,d,i,k_1,k_2)$ has the minimum number of rows and then erasing columns to obtain a matrix with exactly $n$ columns. 
\end{cnstr}

\section{Lower Bounds for \texorpdfstring{$d$-decodable}{d-decodable} Matrices}
So far in the paper we mostly focused on constructions for \decodable[d] matrices which provided upper bounds on the values of $m^*(n,d)$ and $m^*(n,d,k)$, while the exact values was only determined for $d\in\{1,2\}$ and several more special cases in~\autoref{th:basic}. The goal of this section is to close on these gaps and derive lower bounds on the values of  $m^*(n,d)$ and $m^*(n,d,k)$ for several sets of parameters $n,k,$ and $d\ge 3$. 

%For any integers $n,k$ the exact value of $m^*(n,d)$ and $m^*(n,d,k)$ for $d\in\{1,2\}$ is given in \autoref{subsec: basic properties}. In this section we will derive lower bounds on the values of  $m^*(n,d)$ and $m^*(n,d,k)$ for several sets of parameters $n,k$ and $d\ge 3$.

Note that any \decodable[d] matrix $M$ of size $m\times n$ can be considered as a parity-check matrix of some linear code $\Code(M)$ of length $n$ and redundancy $r$. Following this observation, it is possible to derive lower bounds on $m^*(n,d)$ based upon known bounds on linear codes with a prescribed minimum Hamming distance. For all $n$ and $d$, let $K(n,d), r(n,d)$ be the largest dimension, smallest redundancy of any linear code with minimum Hamming distance $d$, respectively, where it holds that $r(n,d) = n - K(n,d)$. The connection between $m^*(n,d)$ and $r(n,d)$ is established in the next theorem, which proved in~\autoref{app:A}.
\begin{theorem}\label{lem: redundancy lower bound}
Let $M$ be a \decodable[d] $m\times n$ matrix, and $\Code(M)$ be the linear code with $M$ as its parity-check matrix. The minimum distance of $\Code(M)$ is at least $d+1$ and $m\ge m^*(n,d)\ge r(n,d+1)$.
\end{theorem}
%\begin{proof}
%Since $M$ is a \decodable[d] matrix, every $d$ columns are linearly independent and thus the minimum Hamming distance of the code $\Code(M)$ is at least $d+1$. 
%Let $\mathsf{k}$ be the dimension of the code $\Code(M)$. By the definition of a parity-check matrix, we have that $M\cdot c^T=0$ if and only if $c\in \Code(M)$. Hence, $\dim(Im(M))=n-k=r$, which implies that the rank of $M$ is $r$ and  $m\ge r$.
%\end{proof}

Using \autoref{lem: redundancy lower bound} we conclude the following lower bounds. The first, second bound uses the sphere packing, Plotkin bound, respectively~\cite{roth_2006}. For proof details see~\autoref{app:A}.
\begin{corollary} \label{corollary_lower_bound}
For all $n$ and $d$ the following properties hold. 
\begin{enumerate}
    \item $m^*(n,d)\ge \log_2 \left( \sum_{i=0}^{\lfloor\frac{d+1}{2}\rfloor}\binom{n}{i}\right) \geq \lfloor \frac{d+1}{2}\rfloor \log_2 \left(\frac{n}{\lfloor\frac{d+1}{2}\rfloor}\right)$
    \item If $d<n<1.5(d+1)$ then $m^*(n,d) = n-1$.
\end{enumerate}
\end{corollary}

The rest of this section is devoted to lower bounds on the value of $m^*(n,d,k)$ for integers $n,d,k$ such that $k\ge 2$ and $d\ge 3$. Here we use another important combinatorial structure in the area of covering codes over fixed weight~\cite{Etzion_Covering}. For positive integers $s,u,v,d$, $K(s,u,v,d)$ refers to the smallest number of length-$n$ binary codewords of weight $u$ such that every binary vector of weight $v$ is at Hamming distance at most $d$ from one of the codewords. This problem was first proposed by Tur\'an himself~\cite{turan} who showed for example that $K(s,2,3,1) = 2\binom{\lfloor\frac{s}{2}\rfloor}{2} +(s\bmod 2)\lfloor\frac{s}{2}\rfloor$. Based on these results we completely solve the case of $d=3, k=2$, which improves upon result derived from the construction of~\cite{Esmaeili09} for this case. Note that the proof provides an explicit construction. Then we present an additional lower bounds on $m^*(m,d,k)$ for any $n\ge 1,d\ge 3$ and $k=d-1$. 

%$\binom{m}{3}$ stopping sets of size $3$. Finding the smallest number of such vectors is equivalent to a well known problem of covering codes over vectors of fixed weight~\cite{Etzion_Covering}. In general, $K(n,u,v,d)$ refers to the smallest number of length-$n$ codewords of weight $u$ such that every vector of weight $v$ is at Hamming distance at most $d$ from one of the codewords. We are interested in the case of $u=2, v=3, d=1$ which was proposed and solved by Tur\'an himself~\cite{??} who showed that $K(n,2,3,1) = 2\binom{\lfloor\frac{n}{2}\rfloor}{2} +(n\bmod 2)\lfloor\frac{n}{2}\rfloor$. Hence, the number of vectors that has to be omitted is $K(m,2,3,1)$ and it is possible to construct an optimal ${m\times (\binom{m}{2}-K(m,2,3,1))}$ \decodable[(3,2)] matrix, which is of size $m\times m^2/4$ when $m$ is even and  $m\times (m^2-1)/4$ when $m$ is odd. 

\begin{theorem}\label{theorem_lower_bound_k=2_d=3}
For any integer $n\ge 1$, we have that  
$$m^*(n,d=3,k=2) =\min\left\{m : \binom{m}{2} - K(m,2,3,1) \geq n\right\} = \ceil{2\sqrt{n}}.$$
\end{theorem}
\begin{proof}
Let $M$ be a \decodable[(3,2)] matrix of size $m\times n$. 
Let $V(m,2)$ be the set of all binary vectors of length $m$ and weight $2$. First note that $|V(m,2)|=\binom{m}{2}$. Next, let us count the number of vector triples $\{u,v,w\}\subseteq V(m,2)$ which are stopping sets. These triples are exactly of the form $u,v,u+v$ for any pair $v,u\in V(m,2)$ which are uniquely characterized by a weight-three vector given by the bitwise OR of the vectors $u,v,u+v$. 
Hence, there are $\binom{m}{3}$ possible stopping sets of size $3$ and each corresponds to a vector of weight three. Since the columns of $M$ must not contain any stopping set of size $3$, we must omit at least one vector from each of these $\binom{m}{3}$ stopping sets. Every vector $w\in V(m,2)$ can resolve $m-2$ out of these stopping sets which are the ones that correspond to vectors of weight 3 and of Hamming distance 1 from $w$. Hence we seek to find the smallest set of vectors $V(m,2)$ that cover all $\binom{m}{3}$ vectors of weight 3. This number is exactly $K(m,2,3,1)$ and hence $n\leq \binom{m}{2} - K(m,2,3,1)$. Given $n$, the value of $m^*(n,d=3,k=2)$ will be the smallest $m$ such that $\binom{m}{2} - K(m,2,3,1) \geq n$ which can be shown to be $2\sqrt{n}$.  
\end{proof}

Similar ideas can be used to derive lower bound on $\minmk{n}{d}{k}$ for any $d \ge 3$ and $k=d-1$. This upper bound is given in the following theorem, and the proof can be found in \autoref{app:A}.

\begin{theorem}\label{theorem_lower_bound_k=d-1}
For any integers $n\ge 1$ and $d\ge 3$ we have that 
%\[ m^*(n,d,k=d-1) \ge (d-1)\sqrt[d-1]{\frac{d}{d-1}n} >  (d-1)\sqrt[d-1]{n}.\]
\[
m^*(n,d,k=d-1) \ge \frac{d-1}{e}\sqrt[d-1]{\frac{nd}{d-1}}.
\]
\end{theorem}

\section{Evaluation }
This section examines practical configurations of the proposed IBLT constructions through numerical examples that complement the analysis  summarized in Table~\ref{table:results}.
%We also examine the performance of the constructions for applications such as error-correction codes.  
To compare our results to the traditional IBLT, we have implemented an IBLT in software.
The table is split into $k$ sub-tables, and items are mapped to cells within each table by the MurmurHash3~\cite{MurmurHash3} hash function.

\subsection{Practical Configurations - LFFZ Size}

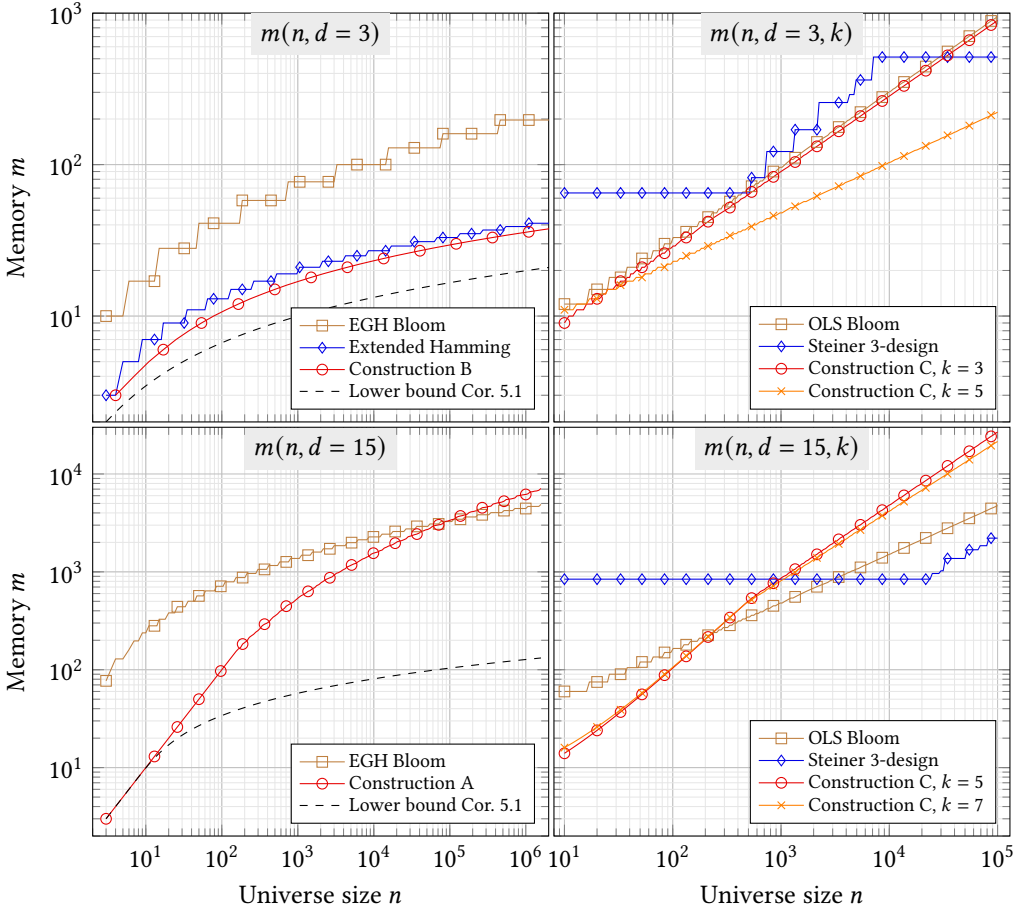
\begin{figure}
\centering

\begin{tikzpicture}

\begin{groupplot}[
group style={
    group size=2 by 2,
    vertical sep=2pt,
    horizontal sep=2pt,
},
title style={
    yshift=-22pt,
    fill=gray!15,
},
xlabel={Universe size $n$},
ylabel={Memory $m$},
ymode=log,
xmode=log,
% xmin=1,
% xmax=7,
% max space between ticks=18,
grid=both,
grid style={line width=.1pt, draw=gray!20},
major grid style={line width=.2pt,draw=gray!50},
height=.5\textwidth,
width=.55\textwidth,
legend cell align={left},
legend style={
    row sep=-1pt,
    nodes={scale=0.75, transform shape},
},
legend columns=1,
legend pos=south east
]

\newcommand{\minxfirst}{2}
\newcommand{\maxxfirst}{2000000}
\newcommand{\minxsecond}{8}
\newcommand{\maxxsecond}{130000}

\newcommand{\minyfirst}{2}
\newcommand{\maxyfirst}{1000}
\newcommand{\minysecond}{2}
\newcommand{\maxysecond}{30000}

\nextgroupplot[
title={$m(n, d=3)$},
xlabel={
% Universe size $n$
},
xticklabels={,,},
ymin=\minyfirst,
ymax=\maxyfirst,
xmin=\minxfirst,
xmax=\maxxfirst,
% legend pos=north west,
]

\addplot[color=brown, mark=square, mark repeat=10]
table [x=n, y=m, col sep=comma]%
{data/nm_EGH_d3.csv};%
\addlegendentry{EGH Bloom}

\addplot[color=blue, mark=diamond, mark repeat=10]
table [x=n, y=m, col sep=comma]%
{data/nm_ex_Hamming_d3.csv};%
\addlegendentry{Extended Hamming}

\addplot[color=red, mark=o, mark repeat=3]
table [x=n, y=m, col sep=comma]%
{data/nm_rec_d3.csv};%
\addlegendentry{Construction B}

\addplot[color=black, dashed, mark repeat=10]
table [x=n, y=m, col sep=comma]%
{data/nm_L61_d3.csv};%
\addlegendentry{Lower bound Cor. 5.1}

\nextgroupplot[
title={$m(n, d=3, k)$},
xlabel={
% Universe size $n$
},
ylabel={},
xticklabels={,,},
yticklabels={,,},
ymin=\minyfirst,
ymax=\maxyfirst,
xmin=\minxsecond,
xmax=\maxxsecond,
]

\addplot[color=brown, mark=square, mark repeat=10]
table [x=n, y=m, col sep=comma]%
{data/nm_OLS_d3.csv};%
\addlegendentry{OLS Bloom}

\addplot[color=blue, mark=diamond, mark repeat=10]
table [x=n, y=m, col sep=comma]%
{data/nm_STNR_d3.csv};%
\addlegendentry{Steiner 3-design}

\addplot[color=red, mark=o, mark repeat=10]
table [x=n, y=m, col sep=comma]%
{data/nm_rec_k3_d3.csv};%
\addlegendentry{Construction C, $k=3$}

\addplot[color=orange, mark=x, mark repeat=10]
table [x=n, y=m, col sep=comma]%
{data/nm_rec_k5_d3.csv};%
\addlegendentry{Construction C, $k=5$}

\nextgroupplot[
title={$m(n, d=15)$},
ymin=\minysecond,
ymax=\maxysecond,
xmin=\minxfirst,
xmax=\maxxfirst,
]

\addplot[color=brown, mark=square, mark repeat=10]
table [x=n, y=m, col sep=comma]%
{data/nm_EGH_d15.csv};%
\addlegendentry{EGH Bloom}

\addplot[color=red, mark=o, mark repeat=10]
table [x=n, y=m, col sep=comma]%
{data/nm_rec_d15.csv};%
\addlegendentry{Construction A}

\addplot[color=black, dashed, mark repeat=10]
table [x=n, y=m, col sep=comma]%
{data/nm_L61_d15.csv};%
\addlegendentry{Lower bound Cor. 5.1}

\nextgroupplot[
title={$m(n, d=15, k)$},
ylabel={},
yticklabels={,,},
ymin=\minysecond,
ymax=\maxysecond,
xmin=\minxsecond,
xmax=\maxxsecond,
]

\addplot[color=brown, mark=square, mark repeat=10]
table [x=n, y=m, col sep=comma]%
{data/nm_OLS_d15.csv};%
\addlegendentry{OLS Bloom}

\addplot[color=blue, mark=diamond, mark repeat=10]
table [x=n, y=m, col sep=comma]%
{data/nm_STNR_d15.csv};%
\addlegendentry{Steiner 3-design}

\addplot[color=red, mark=o, mark repeat=10]
table [x=n, y=m, col sep=comma]%
{data/nm_rec_k5_d15.csv};%
\addlegendentry{Construction C, $k=5$}

\addplot[color=orange, mark=x, mark repeat=10]
table [x=n, y=m, col sep=comma]%
{data/nm_rec_k7_d15.csv};%
\addlegendentry{Construction C, $k=7$}

\end{groupplot}

\end{tikzpicture}

\caption{
The tradeoff between the memory and the universe size allowed by the various constructions for different values of $d$.
The upper charts are for $d=3$, the lower for $d=15$. 
Constructions with arbitrary number of hash functions (\autoref{problem:nd}) are presented on the left column, and with a particular number $k$ (\autoref{problem:ndk}) on the right.
}
\label{fig: lffz size}
\end{figure}

In~\autoref{fig: lffz size} we illustrate for various constructions the memory requirements to allow LFFZ for sets of size at most $d$ selected as a subset of a universe of size $n$. The two top figures refer to $d=3$ such that each presents four constructions, some of them designed specifically for this value of $d$. On the left, constructions with arbitrary number of hash functions (as solutions to \autoref{problem:nd}) and with a particular number $k$ on the right (\autoref{problem:ndk}). We observe that proposed constructions such as Construction A, Construction C and the Extended Hamming (the last two apply specifically for $d=3$), achieve good memory efficiency that is relatively close to the memory lower bound from Corollary~\ref{corollary_lower_bound}.1. The EGH Bloom construction is more expensive but allows to recover each element in the set independently without the necessary to complete the listing of all elements. When the number of hash functions is restricted, we can see in the right the higher memory costs compared to the left figure and in addition how such a constraint can impact a particular construction with the memory cost of Construction D, examined for $k=3$ and $k=5$. The cost for the Steiner 3-design construction, when the universe size is not large, is higher even that of the OLS Bloom construction.
While for larger universe size this construction is preferable.

To allow listing of larger sets, the two bottom figures refer to $d=15$. Here, we see the Bloom-filter based constructions such as  EGH Bloom and OLS Bloom performs well  when the universe size increases. The universe size $n$ has a small impact on the memory for the Steiner 3-design constructions that fits well large $n$ values. We also see an increasing gap between the constructions and the lower bound, leaving hope for additional constructions that are efficient especially for large $d$ values.

\subsection{Failure Probability beyond the LFFZ}
In~\autoref{fig:intro}, we 
examine the probability of successful listing of the traditional IBLT vs. our suggested IBLT with LFFZ. 
In the top figure we considered a universe size $n=|U|=25$ and memory of size $m=15$, while in the bottom figure the universe is of size $n=|U|=381$ and $m=64$. 
In both cases, elements were drawn randomly from $U$ and inserted into the IBLT,IBLT with LFFZ and then a listing was performed. We repeat this process $10^6$ times for each list size and the presented success probability is the mean of all this tests.    

It can be observed that the success probability is a function of the table length and number of elements in the table.
Both figures show that in the traditional IBLT, listing can fail even with a set of two elements, while for the IBLT with LFFZ listing is always successful for any number of elements up to some known parameter $d$, allowed by the table length and the universe size. Furthermore, it can be noticed that the IBLT with LFFZ outperforms the traditional IBLT even when the number of elements in the table exceeds the value $d$.  

\subsection{Memory Consideration}
As explained in \autoref{subsec: data structures}, in a traditional IBLT, mapping an element in the universe to cells is usually done using $k$ hash functions.
Hence, the additional memory required for the mapping is minor and usually requires $O(k)$ additional memory.
In our model, the mapping is done using an $m \times n$ matrix, and while with trivial implementation the matrix induces high memory requirements for large values of $n$, it can be significantly reduced in practice as will be discussed in this section.

First, note that the mapping matrix is binary and can be stored using $mn$ bits. However, in some of the constructions, most elements are zeros, and these matrices can be stored in a sparse form. This holds in particular for solutions for \autoref{problem:ndk}, where we require a fixed number of ones per column. For instance, the mapping matrix for the Array~codes given in \autoref{section_linear_codes_constructions} contains up to 4 ones in every column. Therefore, each column can be represented by the 4 locations, and the whole matrix can be stored using at most $4n\log_2{m}$ bits.

A different approach is to trade the mapping memory with mapping computation.
For many constructions, instead of storing the whole mapping matrix in memory, we can generate a single column at a time for each element needed to be mapped. Taking the mapping matrix of the EGH Bloom filter for example (\autoref{section_bloom_constructions}), the $i$-th column of the matrix can be easily generated by concatenating $i\bmod{q_j}$ for a list of the smallest primes $q_j$. Thus, by storing the required list of primes, one can reconstruct a column on demand without holding the full matrix.
Similarly, all of our recursive constructions (\autoref{section_recursive_constructions}) are suitable for on demand column generation. In \autoref{const: improved 3-decodable}, we can construct a single column by recursively computing the locations within the sub-matrices using the equations in the proof of \autoref{th:d3construction}.
The same can be done with \autoref{const: improved d-decodable} and \autoref{const: basic (d,k)-decodable}, with a small modification;
we can fix the parameters $i,k_1,k_2$ and derive similar equations as in \autoref{const: improved 3-decodable} for fast column computation without additional memory.

\section{Conclusions and Future Work}
Invertible Bloom Lookup Table (IBLT) is a major building block in several networking applications but its potential for listing failures 
implies high overhead cost. We suggest to rely on coding techniques to propose a new approach to its design with constructions of guaranteed listing when sets satisfy an upper bound on their size. We focus on listing that relies on  peeling elements in pure cells. A natural open question for future work refers to the design of IBLTs that allow listing, not necessarily through the peeling procedure such as through solving a system of equations.

% \begin{acks}
% \end{acks}

% \bibliographystyle{ACM-Reference-Format}
\bibliographystyle{unsrt}
\bibliography{IBLTcodes}

\begin{thebibliography}{10}

\bibitem{IBLT}
Michael~T. Goodrich and Michael Mitzenmacher.
\newblock Invertible bloom lookup tables.
\newblock In {\em Allerton Conference on Communication, Control, and
  Computing}, 2011.

\bibitem{li2016lossradar}
Yuliang Li, Rui Miao, Changhoon Kim, and Minlan Yu.
\newblock {Lossradar}: {F}ast detection of lost packets in data center
  networks.
\newblock In {\em ACM International on Conference on emerging Networking
  EXperiments and Technologies (CoNext)}, 2016.

\bibitem{li2016flowradar}
Yuliang Li, Rui Miao, Changhoon Kim, and Minlan Yu.
\newblock Flowradar: A better netflow for data centers.
\newblock In {\em USENIX Symposium on Networked Systems Design and
  Implementation (NSDI)}, 2016.

\bibitem{BiffCodes}
Michael Mitzenmacher and George Varghese.
\newblock Biff ({B}loom filter) codes: Fast error correction for large data
  sets.
\newblock In {\em {IEEE} International Symposium on Information Theory
  {(ISIT)}}, 2012.

\bibitem{EppsteinGUV11}
David Eppstein, Michael~T. Goodrich, Frank Uyeda, and George Varghese.
\newblock What's the difference?: {E}fficient set reconciliation without prior
  context.
\newblock In {\em {ACM} {SIGCOMM}}, 2011.

\bibitem{multiparty}
Michael Mitzenmacher and Rasmus Pagh.
\newblock Simple multi-party set reconciliation.
\newblock {\em Distributed Comput.}, 31(6):441--453, 2018.

\bibitem{Graphene19}
A.~Pinar Ozisik, Gavin Andresen, Brian~Neil Levine, Darren Tapp, George
  Bissias, and Sunny Katkuri.
\newblock Graphene: {E}fficient interactive set reconciliation applied to
  blockchain propagation.
\newblock In {\em {ACM} {SIGCOMM}}, 2019.

\bibitem{Bloom}
Burton~H. Bloom.
\newblock Space/time trade-offs in hash coding with allowable errors.
\newblock {\em Commun. ACM}, 13(7):422--426, 1970.

\bibitem{stoppingSets}
Changyan Di, David Proietti, I.~Emre Telatar, Thomas~J. Richardson, and
  R{\"{u}}diger~L. Urbanke.
\newblock Finite-length analysis of low-density parity-check codes on the
  binary erasure channel.
\newblock {\em IEEE Transactions on Information theory}, 48(6):1570--1579,
  2002.

\bibitem{schwartz2006ss}
Moshe Schwartz and Alexander Vardy.
\newblock On the stopping distance and the stopping redundancy of codes.
\newblock {\em IEEE transactions on information theory}, 52(3):922--932, 2006.

\bibitem{EGH}
S{\'{a}}ndor~Z. Kiss, {\'{E}}va Hosszu, J{\'{a}}nos Tapolcai, Lajos
  R{\'{o}}nyai, and Ori Rottenstreich.
\newblock Bloom filter with a false positive free zone.
\newblock In {\em {IEEE} {INFOCOM}}, 2018.

\bibitem{FPFZ}
Ori Rottenstreich, Pedro Reviriego, Ely Porat, and S.~Muthukrishnan.
\newblock Constructions and applications for accurate counting of the {B}loom
  filter false positive free zone.
\newblock In {\em {ACM} Symposium on SDN Research {(SOSR)}}, 2020.

\bibitem{etzion2006stopping}
Tuvi Etzion.
\newblock On the stopping redundancy of {Reed}--{Muller} codes.
\newblock {\em IEEE transactions on information theory}, 52(11):4867--4879,
  2006.

\bibitem{Esmaeili09}
Morteza Esmaeili and Mohammad~Javad Amoshahy.
\newblock On the stopping distance of array code parity-check matrices.
\newblock {\em IEEE Transactions on Information Theory}, 55(8):3488--3493,
  2009.

\bibitem{laendner2007ldpc}
Stefan Laendner and Olgica Milenkovic.
\newblock {LDPC} codes based on {L}atin squares: Cycle structure, stopping set,
  and trapping set analysis.
\newblock {\em IEEE Transactions on Communications}, 55(2):303--312, 2007.

\bibitem{roth_2006}
Ron Roth.
\newblock {\em Introduction to Coding Theory}.
\newblock Cambridge University Press, 2006.

\bibitem{colbourn1999triple}
Charles~J Colbourn and Alexander Rosa.
\newblock {\em Triple systems}.
\newblock Oxford University Press, 1999.

\bibitem{blanchard1995construction}
John~L Blanchard.
\newblock A construction for steiner 3-designs.
\newblock {\em Journal of Combinatorial Theory, Series A}, 71(1):60--66, 1995.

\bibitem{CoveringArrays}
Jim Lawrence, Raghu Kacker, Yu~Lei, D.~Richard Kuhn, and Michael~A. Forbes.
\newblock A survey of binary covering arrays.
\newblock {\em Electron. J. Comb.}, 18(1), 2011.

\bibitem{yugawa2014finite}
Daichi Yugawa and Tadashi Wadayama.
\newblock Finite length analysis on listing failure probability of invertible
  {Bloom} lookup tables.
\newblock {\em IEICE Transactions on Fundamentals of Electronics,
  Communications and Computer Sciences}, 97-A(12):2309--2316, 2014.

\bibitem{kubjas2020failure}
Ivo Kubjas and Vitaly Skachek.
\newblock Failure probability analysis for partial extraction from invertible
  {Bloom} filters.
\newblock {\em arXiv preprint arXiv:2008.00879}, 2020.

\bibitem{Lazaro_irregular}
Francisco L{\'{a}}zaro and Bal{\'{a}}zs Matuz.
\newblock Irregular invertible {Bloom} look-up tables.
\newblock {\em CoRR}, abs/2107.02573, 2021.

\bibitem{jiang2017parallel}
Jiayang Jiang, Michael Mitzenmacher, and Justin Thaler.
\newblock Parallel peeling algorithms.
\newblock {\em ACM Transactions on Parallel Computing (TOPC)}, 3(1):1--27,
  2017.

\bibitem{superimposedcodes}
William~H. Kautz and Richard~C. Singleton.
\newblock Nonrandom binary superimposed codes.
\newblock {\em IEEE Transactions on Information theory}, 10(4):363--377, 1964.

\bibitem{eppstein2007improved}
David Eppstein, Michael~T Goodrich, and Daniel~S Hirschberg.
\newblock Improved combinatorial group testing algorithms for real-world
  problem sizes.
\newblock {\em SIAM Journal on Computing}, 36(5):1360--1375, 2007.

\bibitem{Esmaeili_LDPC}
M.~Esmaeili, M.~H. Tadayon, and T.~A. Gulliver.
\newblock More on the stopping and minimum distances of array codes.
\newblock {\em IEEE Transactions on Communications}, 59(3):750--757, 2011.

\bibitem{rosnes2014minimum}
Eirik Rosnes, Marcel~Adrian Ambroze, and Martin Tomlinson.
\newblock On the minimum/stopping distance of array low-density parity-check
  codes.
\newblock {\em IEEE transactions on information theory}, 60(9):5204--5214,
  2014.

\bibitem{linial2022bounds}
Nati Linial and Idan Orzech.
\newblock Bounds on unique-neighbor codes.
\newblock {\em arXiv preprint arXiv:2203.10330}, 2022.

\bibitem{kashyap2003stopping}
Navin Kashyap and Alexander Vardy.
\newblock Stopping sets in codes from designs.
\newblock In {\em IEEE International Symposium on Information Theory (ISIT)},
  2003.

\bibitem{Etzion_Covering}
Tuvi Etzion, Victor Wei, and Zhen Zhang.
\newblock Bounds on the sizes of constant weight covering codes.
\newblock {\em Des. Codes Cryptography}, 5:217--239, 05 1995.

\bibitem{turan}
Tur\'an. Peter.
\newblock On the theory of graphs.
\newblock {\em Colloq. Math.}, 3:146--163, 1954.

\bibitem{MurmurHash3}
Austin Appleby.
\newblock Murmurhash3, 2016.
\newblock
  https://github.com/aappleby/smhasher/blob/\\master/src/MurmurHash3.cpp.

\end{thebibliography}

\appendix

\section{Appendix}\label{app:A}

\paragraph{Proof of \autoref{th:basic}}
Let $n,k$ and $d\ge 3$ be integers.
First note that since any \decodable[(d,k)] matrix is also a \decodable[d] matrix and all columns are different it holds that $m^*(n,k,d)\ge m^*(n,d) \ge \ceil{\log_2\del{n+1}}$.
In addition, any set of $d$ binary vectors of length $\ell<d$ must be linearly dependent, and in particular, contains a stopping set.
Hence, $m^*(n,d,k) \ge m^*(n,d)\ge d$. $m^*(n,d,k) \ge k$ since the weight of every column is $k$.
The second claim follows since, by definition, any \decodable[d] matrix is also a \decodable[(d-1)] matrix.   
The third claim is established by the $n\times n$ identity matrix when $d=n$ and for $d<n$, it follows from the $(n-1)\times (n-1)$ identity matrix with another all-one column.
The fourth claim uses the same $n\times n$ identity matrix while adding more $k-1$ rows of all-one vectors.
Since for any $\ell\ge 1$ there are exactly $\ell$ distinct binary vectors of length $\ell$ and weight one, any \decodable[(d,1)] matrix with $n$ columns must have at least $n$ rows, i.e., $m^*(n,d,k=1) \ge n$. 
%The fifth claim follows from first and third claims. 
The fifth claim is established by constructing a \decodablek{d}{k+1} from a \decodablek{d}{k} while adding an all-one row. 
%Since the $n\times n$ identity matrix is \decodable[(d,1)] for any integer $d$, this bound is tight. The third claim is established similarly by the $n\times n$ identity matrix and the fourth claim follows from the first and third claims. 

\paragraph{Proof of \autoref{th:linear codes to d-decodable}}
By the definition of stopping redundancy, there exists a parity-check matrix $H$ for the code $\Code$ with $\rho$ rows, $n$ columns and stopping distance $d$. Since the stopping distance of $H$ is $d$, any set of at most $d-1$ columns is not a stopping set and thus $H$ is the required \decodable[(d-1)] matrix of size $\rho\times n$.
%Let $H$ be a parity-check matrix for the code $\Code$ with $\rho=\rho\del{\Code}$ rows and $n$ columns. We define the binary matrix $M$ of size $\rho \times n$ as follows. $$ M_{i,j} = 1 \iff H_{i,j} \ne 0,$$ i.e. $M$ is the binary matrix with the same dimensions as $H$, such that any non-zero element of $H$ is replaced by $1$ in $M$. For $\ell\le d-1$, let $\cbr{m_{j_1},m_{j_2},\ldots m_{j_\ell}}$ be a set of $\ell$ columns of $M$. To prove that $M$ is \decodable[(d-1)] it is sufficient to show that the $\rho\times \ell$ sub-matrix $M'$ which consists from these columns contains a row of weight one.
%Assume to the contrary that there is no such row, and consider the sub-matrix $H'$ that is obtained from the corresponding $\ell$ columns of $H$, $\cbr{h_{j_1},\ldots,h_{j_\ell}}$. It is readily verified that for any $1\le i\le \rho$ the weight of the $i$-th row of $M'$ and the weight of the $i$-th row of  $H'$ are equal. Hence, $H'$ forms a stopping set of size $\ell\le d-1$ in $H$ which is a contradiction since the stopping distance of $H$ is $d$.    

\paragraph{Proof of Corollary~\ref{cor: Steiner and IBLT}}
Before we prove the corollary, we present a well known expression for the number of columns in the incidence matrix of a Steiner system.
For a Steiner system $S=S(t,k,m)$, the number of blocks in $S$ is equal to $\binom{m}{t}\big\slash\binom{k}{t}$.
The results follow from the latter property, \autoref{the: stopping set in an incidence structre}, and the following two families of Steiner systems.
\begin{enumerate}
    \item A Steiner system $S=S(2,3,m)$, is called a \emph{Steiner triple system of order $m$}, or $STS(m)$ and it is known that an $STS(m)$ exists if and only if $m(\bmod{6})\in \{1,3\}$~\cite{colbourn1999triple}. Hence, for any $m$ such that $m(\bmod{6})\in \{1,3\}$, the $STS(m)$ forms a \decodable[(3,3)] matrix with $n=\binom{m}{2}\slash 3=\frac{m(m-1)}{6}$ and thus 
    $$m^*(n,d = 3,k=3) \le \left\lceil\frac{1+\sqrt{1+24n}}{2}\right\rceil+3=\Theta(\sqrt{n} ) .$$
    \item There are known constructions of Steiner system $S=S(3,q+1,q^\alpha +1)$ for any prime power $q$ and any integer $\alpha\ge 2$~\cite{blanchard1995construction}. Any such a system forms a \decodable[(\lceil\frac{q+1}{2}\rceil,q+1)] matrix with, $m=q^\alpha+1$ and $n = \binom{m}{3}\big\slash\binom{q+1}{3} = \frac{m(m-1)(m-2)}{(q+1)q(q-1)} = \frac{m^3-3m^2+2m}{q(q^2-1)}$ and thus it can be derived that
    \[
   m^*\left(n,d=\left\lceil\frac{q+1}{2}\right\rceil,k=q+1 \right)\le O(d^2\sqrt[3]{n}).
    \]
\end{enumerate}

% \paragraph{Proof of \autoref{th: basic d-decodable}}
% It can be easily verified that for any $d\ge 3$, the matrices $M^n(d)$, $n< 1.5(d+1)$ are \decodable[d]. The proof that $M^n(d)$ for $n\ge 1.5(d+1)$ is \decodable[d] can be done similarly to the proof of \autoref{th: recursive basic}.

\paragraph{Sketch proof of \autoref{theorem_construction_A}}
For $d=3$ and by setting $i=2$ in \autoref{const: improved d-decodable}, one can derive the following relation on the number of rows in the matrices $M(n,3,2)$: $\Rows{M(n,3,2)} = \Rows{M(\lceil n/2\rceil,3,2)} + 2$. Since $\Rows{M(4,3,2)} = 3$, it holds that $\Rows{M(n,3,2)} = 2\lceil \log_2 n\rceil - 1$ and thus $m^*(n,3)\le 2\ceil{\log_2 n} -1$. The proof for arbitrary values of $d$ continues by a simple induction on $d$. 

\paragraph{Proof of \autoref{lem: i=3}} Let $f(m)$ denote the number of columns in the matrix $M_m$ obtained by the construction. It holds that 
$$\argmax_{i \in \sbr{2,m-3}}{\del{i\cdot \Cols{M_{m-i}}}} = 
\argmax_{i \in \sbr{2,m-3}}{\del{i\cdot f(m-i)}}.
$$
The proof is done by induction over $m$. It can be verified using computer search that $f(6)=17$ and $f(9)=54$, and hence $f(9)=3+3f(6)$ and $i=3$. Assume that the claim is correct for any $9\le m'<m$. We give the proof assuming $m-i$ is divisible by $3$ while similar proof holds also for $m$ such that $(m-i) \bmod 3\not\equiv 0$.
By the induction assumption we have that
\begin{align*}
f(m-i) & = 3 + 3f(m-i-3) = 3 + 9 + 9f(m-i-6)\\& = \cdots = \sum_{j=1}^{(m-i-9)/3}3^j + 3^{(m-i-9)/3}\cdot f(9)\\
& = \frac{3}{2}\left(3^{(m-i-9)/3}-1\right) + 3^{(m-i-9)/3}\cdot 54\\
& = 55.5\cdot 3^{(m-i-9)/3} - 3/2.
\end{align*}
It can be verified using derivation that the expression 
\begin{align*}
i\cdot f(m-i) = i\cdot 55.5\cdot 3^{(m-i-9)/3} - \frac{3i}{2}
\end{align*}
is maximized over the integers whenever $i=3$.

\paragraph{Proof of \autoref{th:d3construction}}
The number of columns in $M_3$ and $M_4$ are $3$ and $7$, respectively. For $5\le m \le 9$, the number of columns in $M_m$ can be found using a computer search and the results are as follows: $\Cols{M_5}=11$,
$\Cols{M_6}=17$,
$\Cols{M_7}=25$,
$\Cols{M_8}=37$,
$\Cols{M_9}=54$. 
For $m>9$, \autoref{lem: i=3} states that $i=3$ and hence 
$$
\Cols{M_m} = 3 + 3\Cols{M_{m-3}},
$$
or equivalently,
\begin{align*}
\Cols{M_m}&=
\begin{cases}
\sum_{j=1}^{(m-9)/3}3^j + 3^{(m-9)/3}\cdot \Cols{M_9}  ,& m\bmod 3 \equiv 0 \\
\sum_{j=1}^{(m-7)/3}3^j + 3^{(m-7)/3}\cdot \Cols{M_7} , & m\bmod 3 \equiv 1 \\
\sum_{j=1}^{(m-8)/3}3^j + 3^{(m-8)/3}\cdot \Cols{M_8} , & m\bmod 3 \equiv 2
\end{cases}
\\ 
& = \begin{cases}
\frac{1}{18}\del{37\cdot 3^{m/3}-27     }, & m\bmod 3 \equiv 0 \\
\frac{1}{2} \del{53\cdot 3^{(m-7)/3} - 3}, & m\bmod 3 \equiv 1 \\
\frac{1}{2} \del{77\cdot 3^{(m-8)/3} - 3}, & m\bmod 3 \equiv 2.
\end{cases}
\end{align*}

By extracting $m$ we have that
\begin{align*}
m =
\begin{cases}
\frac{3}{\log_2 3}\cdot \log_2{\del{\frac{18}{37}\Cols{M_m}+ \frac{27}{37}}}     , & m\bmod 3 \equiv 0 \\
\frac{3}{\log_2 3}\cdot \log_2{\del{\frac{2}{53} \Cols{M_m}+ \frac{3}{53} }} + 7 , & m\bmod 3 \equiv 1\\
\frac{3}{\log_2 3}\cdot \log_2{\del{\frac{2}{77} \Cols{M_m}+ \frac{3}{77} }} + 8 , & m\bmod 3 \equiv 2,
\end{cases}
\end{align*}
and for $m\ge 3$, we have $\Cols{M_m}\ge 3$ and  ${m\le \frac{3}{\log_2 3}\cdot \log_2{\del{\Cols{M_m}}}}$. Thus, $\minm{n}{3} \le \recresult$.

\paragraph{Proof of \autoref{lem: redundancy lower bound}}
Since $M$ is a \decodable[d] matrix, every $d$ columns are linearly independent and thus the minimum Hamming distance of the code $\Code(M)$ is at least $d+1$. 

\paragraph{Proof of Corollary~\ref{corollary_lower_bound}}
The first bound is a direct result of~\autoref{lem: redundancy lower bound} together the sphere packing bound which asserts that $A(n,d+1) \leq \frac{2^n}{\sum_{i=0}^{\lfloor\frac{d+1}{2}\rfloor}\binom{n}{i}}$. Thus  
$r(n,d+1) \geq \log_2 \left( \sum_{i=0}^{\lfloor\frac{d+1}{2}\rfloor}\binom{n}{i}\right) \geq \lfloor \frac{d+1}{2}\rfloor \log_2\left(\frac{n}{\lfloor\frac{d+1}{2}\rfloor}\right)$, where the last bound follows from $\sum_{i=0}^t\binom{n}{i} \geq \binom{n}{t}\geq \frac{n^t}{t^t}$.

The second bound uses the Plotkin bound which claims that $K(n,d+1)\leq \log_2\left(2\left\lfloor\frac{d+1}{2d+2-n}\right\rfloor\right)$. Hence, if $n<1.5(d+1)$ then $\left\lfloor\frac{d+1}{2d+2-n}\right\rfloor = 1$ and then $K(n,d+1) \geq 1$, i.e., $m^*(n,d)\geq r(n,d+1) \geq n-1$. Together with~\autoref{th:basic}(3), strict equality holds.

\paragraph{Proof of \autoref{theorem_lower_bound_k=d-1}}
Similar to the proof of \autoref{theorem_lower_bound_k=2_d=3}, let $V(k=d-1,m)$ be the set of all binary columns of length $m$ and weight $d-1$. For any set of $d$ indices $1\le i_1<\cdots<i_d\le m$, there exists a unique set of $d$ columns $\{v_1,\ldots v_d\}\subset V(d-1,m)$ such that $$\{i_1,\ldots,i_d\}=\bigcup_{1\le j\le d}\mathsf{supp}(v_{j}).$$ Any such set is a stopping set of size $d$ and all the sets are unique. Hence, there are at least $\binom{m}{d}$ stopping sets of size $d$ within $V(d-1,m)$. Since any \decodable[(d,d-1)] matrix must not contain any stopping set of size $d$, we must omit at least one vector from each of the $\binom{m}{d}$ stopping sets of size $d$. Hence, the minimum number of vectors that we need to remove to solve all of these stopping sets is $K(m,d-1,d,1)$. Thus, 
$$n\leq \binom{m}{d-1} - K(m,d-1,d,1),$$
and since $K(m,d-1,d,1)\geq \frac{\binom{m}{d}}{m-d+1}$, we also get that 
$$n \leq \binom{m}{d-1} - \frac{\binom{m}{d}}{m-d+1} = \binom{m}{d-1}\cdot \left(1- \frac{1}{d}\right).$$
Therefore, 
\begin{align*}
m^*(n,d-1,d) & \geq \min\left\{ m: \binom{m}{d-1} - K(m,d-1,d,1)\right\} & \\
& \geq \min\left\{ m: \binom{m}{d-1}\cdot \left(1- \frac{1}{d}\right)\right\}. & 
\end{align*}
%Any column vector $v\in V(d-1,m)$ belongs to exactly $m-d+1$ such stopping sets and hence we must omit at least $\binom{m}{d}\cdot \frac{1}{m-d+1}$ columns. Hence 
%\begin{align*}
%    n & < |V(d-1,m)| - \binom{m}{d}\cdot \frac{1}{m-d+1} \\ & =
%     \binom{m}{d-1} - \binom{m}{d}\cdot \frac{1}{m-d+1} \\ & = 
%      \binom{m}{d-1}\cdot \left(1- \frac{1}{d}\right),
%\end{align*}
Lastly, from $\binom{m}{d-1} \le \left(\frac{em}{d-1}\right)^{d-1}$,
we get that $m\ge \frac{d-1}{e}\sqrt[d-1]{\frac{nd}{d-1}}$.

\end{document}